\documentclass[fleqn,usenatbib,useAMS]{mnras}

\usepackage{newtxtext,newtxmath}
\usepackage{braket}
\usepackage{natbib}
\usepackage{booktabs}
\usepackage{multirow}
\usepackage{ulem}
\usepackage[dvipsnames]{xcolor}

\usepackage[T1]{fontenc}

\DeclareRobustCommand{\VAN}[3]{#2}
\let\VANthebibliography\thebibliography
\def\thebibliography{\DeclareRobustCommand{\VAN}[3]{##3}\VANthebibliography}

\usepackage{graphicx}	
\usepackage{amsmath}	
\usepackage{changes}                                         
\usepackage{lineno}

\title[A holistic model of magnetic braking -- II]{Towards a holistic magnetic braking model -- II: explaining several long-term internal- and surface-spin properties of solar-like stars and the Sun}
\author[Sarkar et al.]{
Arnab Sarkar$^{1}$\thanks{E-mail: as3158@cam.ac.uk},
Patrick Eggenberger$^{2}$,
Lev Yungelson$^{3}$
and  Christopher A. Tout$^{1}$\thanks{E-mail: cat@ast.cam.ac.uk}
\\
\\
$^{1}$Institute of Astronomy, The Observatories, Madingley Road, Cambridge CB3 OHA, UK
\\
$^{2}$ Département d’Astronomie, Université de Genève, Versoix, Switzerland 
\\
$^{3}$Institute of Astronomy of the Russian Academy of Sciences, 48 Pyatnitskaya Str.,119017 Moscow, Russia
}

\date{Accepted XXX. Received YYY; in original form ZZZ}

\pubyear{2023}

\begin{document}
\label{firstpage}
\pagerange{\pageref{firstpage}--\pageref{lastpage}}
\maketitle

\begin{abstract}
We extend our model of magnetic braking (MB), \textcolor{black}{driven by an $\alpha-\Omega$ dynamo mechanism}, from fully convective M-dwarfs (FCMDs) to explain the surface and internal spin $P_\mathrm{spin}$ evolution of partly convective dwarfs (PCDs) starting from the disc-dispersal stage to the main-sequence turnoff. In our model, the spin of the core is governed by shear at the core-envelope boundary while the spin of the envelope is governed by MB and shear. We show that (1) the most massive FCMDs experience a stronger spin-down than PCDs and less massive FCMDs, (2) the stalled spin-down and enhanced activity of K-dwarfs and the pileup of G-dwarfs older than a few Gyr are stellar-structure- and MB-dependent, and weakly dependent on core-envelope coupling effects, (3) our expression of the core-envelope convergence time-scale $\tau_\mathrm{converge}(M_\ast,\,P_\mathrm{spin})$ between a few 10 to 100~Myr strongly depends on stellar structure but weakly on MB strength and shear, such that fast and massive rotators achieve corotation earlier, (4) our estimates of the surface magnetic fields are in general agreement with observations and our wind mass loss evolution explains the weak winds from the solar analog $\pi^1$ UMa and (5) with our model the massive young Sun hypothesis as a solution to the faint young Sun problem can likely be ruled out, because the maximum mass lost by winds from our Sun with our model is about an order of magnitude smaller than required to solve the problem. 

\end{abstract}


\section{Introduction} 
\defcitealias{SYT}{SYT~I}

Magnetic braking (MB) is a mechanism by which stellar angular momentum is lost \citep{1962AnAp...25...18S}. It operates in many single and binary stars and in many cases drives the long-term spin and orbital properties of a system. Accurate modelling of the MB torque and its evolution is especially crucial for the studies of low-mass stars and cataclysmic variables \citep{2003ApJ...586..464B, 2003cvs..book.....W}. 

\cite{1972ApJ...171..565S} found that the equatorial rotation velocity of single G-dwarfs scales with their age $t$ as $t^{-0.5}$. The consequent torque scales with spin period $P_\mathrm{spin}$ as $P_\mathrm{spin}^{-3}$ \citep{1981A&A...100L...7V}. However, it was soon evident that the MB torque has a weaker dependence on the rotational angular velocity $\Omega$ in rapidly spinning stars \citep{1988ApJ...333..236K, 2011ApJ...743...48W}. This has been incorporated in various MB models \citep{Sills2000, Matt2015}. Observations of isolated low-mass stars in open clusters (OCs) of known ages also revealed a bimodality in their rotation rates: the same OC contains a population of fast as well as slow rotators \citep{2003ApJ...586..464B}. Efforts have been made to explain this with mechanisms such as stochastic changes in the MB torque \citep{Brown2014} or multipolar effects in the stellar magnetic fields \citep{Garraffo2018}.

Studying partly convective dwarfs (PCDs, $0.35\lesssim M_\ast/\,M_\odot\lesssim1.3$) poses an additional problem. While the evolution of $P_\mathrm{spin}$ of fully convective M-dwarfs (FCMDs, $M_\ast\lesssim0.35\,M_\odot$) can be well modelled assuming the whole star rotates as a solid body (SB), modelling PCDs as SBs is inconsistent with observations \citep{Denissenkov2010}. This is because PCDs possess a radiative core and a convective envelope. The radiative interior is not as efficient as the convective envelope at transporting angular momentum so the interior of the star can decouple, breaking down the SB rotation for such stars. So, modelling PCDs requires a mechanism for the transport of internal angular momentum within the star. This has been done in detailed stellar evolution codes that incorporate shellular rotation \citep{1992A&A...265..115Z} in a star with angular momentum transport via the Tayler-Spruit dynamo \citep{2002A&A...381..923S,2022NatAs...6..788E} by \cite{Denissenkov2007} and \cite{Denissenkov2010} and with a simpler two-zone model by \cite{1991ApJ...376..204M}, who used characteristic core-envelope coupling time-scales \citep{Denissenkov2010}. However, both of these approaches depend on two uncertain and interrelated mechanisms, MB and internal angular momentum transport (in the form of the coupling time-scale in the latter approach). The problem is two-fold: not only do we require the surface rotation rates of our models to agree with robust observations \citep{2021ApJS..257...46G, Pass2022}, but we also require that the core and the envelope corotate within Gyr time-scales as shown by the internal rotation of the Sun (\citealt{Thompson2003}) and suggested by asteroseismic data for stars with $M_\ast\gtrsim 0.8M_\odot$ \citep{Btrisey2023}. Several aspects of the spin evolution of PCDs have been addressed using a semi-empirical approach \citep{SadeghiArdestani2017} or a holistic framework combining stellar activity and spindown \citep{2016MNRAS.458.1548B}.

In a series of papers, we have illustrated that robust modelling of MB can help explain the observational properties of many single- and binary-star systems (\citealt{2022MNRAS.513.4169S} for cataclysmic variables, \citealt{2023MNRAS.519.2567S} and \citealt{Sarkar2023} for AM Canum Venaticorum stars). Recently \citet[hereinafter \citetalias{SYT}]{SYT} used an $\alpha-\Omega$ dynamo mechanism of MB to explain the spin evolution of FCMDs. 
In this paper, we extend our analysis to holistically model partly convective low-mass stars and address several theoretical and observational constraints on their evolution. 

In Section~\ref{sec:math} we provide the details of our model of MB and angular momentum transport. In Section~\ref{sec:spind} we study the behaviour of our model in PCDs of various masses and the dependence of the spin evolution of the core and the envelope on uncertain parameters in our MB and angular momentum transport mechanism. We present our results and discuss their implications in Section~\ref{sec:results}. We conclude in Section~\ref{sec:conclusions}.

\section{Magnetic braking and internal transport of angular momentum}
\label{sec:math}

\begin{figure*}
\centering
\includegraphics[width=0.85\textwidth]{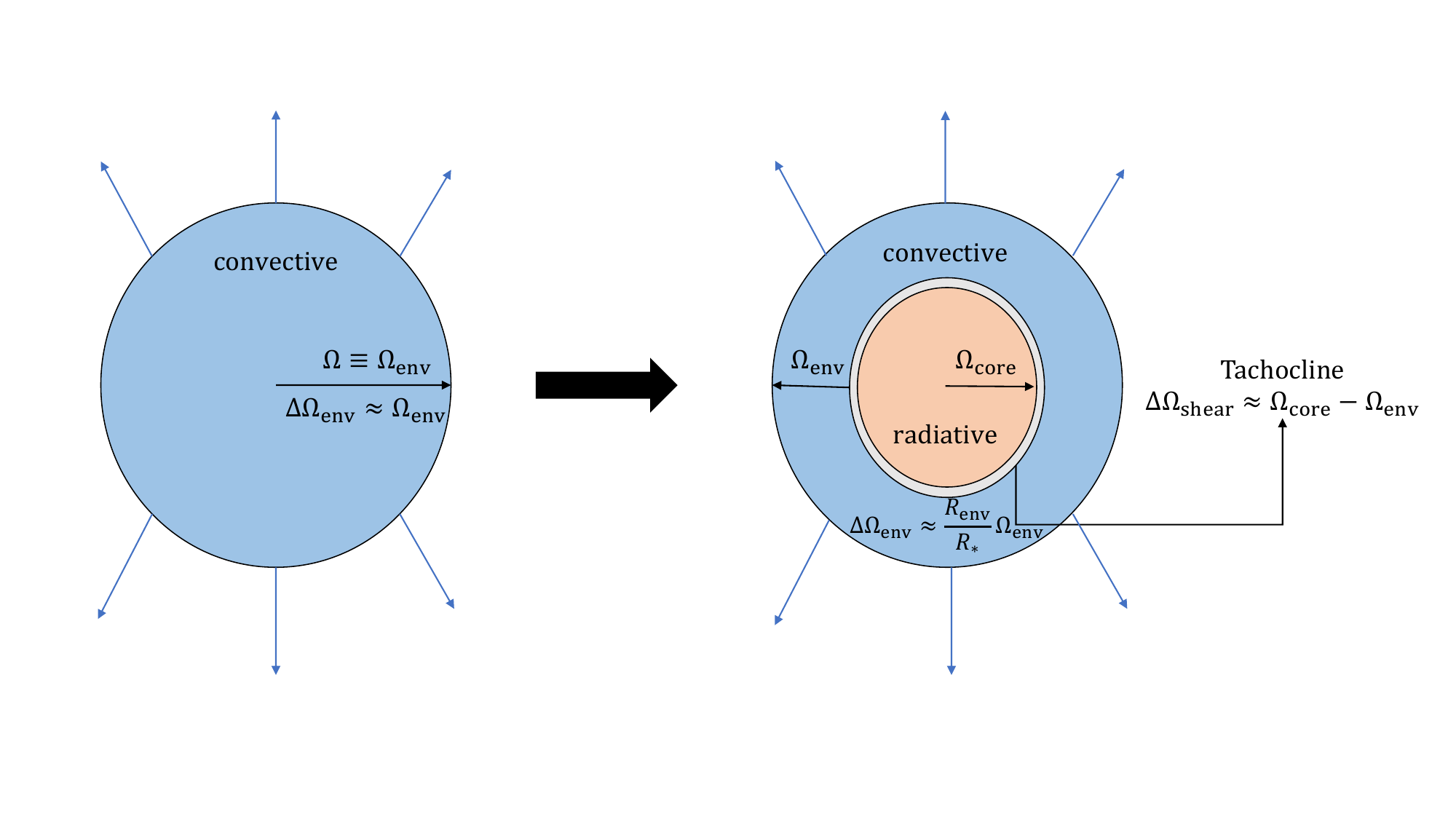}
\caption{A schematic diagram of the various parameters with which we model the spin of the core and envelope of a fully convective pre-main-sequence star (left) that evolves to a partly convective main-sequence low-mass star (right). The blue lines emerging out of the surface of the star denote MB due to our $\alpha-\Omega$ dynamo mechanism (Section~\ref{subs:conv_dynamo}). The schematic on the left also describes the evolution of fully convective M-dwarfs.  }
\label{fig:cartoon}
\end{figure*}
A schematic diagram illustrating the core and envelope properties of PCDs in our model is shown in Fig.~\ref{fig:cartoon}. In our model, MB in the star is caused by an $\alpha-\Omega$ dynamo operating in the convective zone. We model this with the formalism of \citet[hereinafter TP]{1992MNRAS.256..269T} for contracting pre-main sequence (PMS) stars. We use a two-zone model with a uniformly rotating radiative core and convective envelope \citep{1991ApJ...376..204M} but employ a mechanism of shear according to \cite{Zangrilli1997}.


\subsection{The convective dynamo}
\label{subs:conv_dynamo}
The basic formulation of our $\alpha-\Omega$ dynamo in the convective envelope is the same as in \citetalias{SYT}, albeit with a few updates. We urge the reader to refer to this paper and only highlight here the changes made to the model for a partly convective envelope. Hereinafter the subscript env denotes the convective envelope (star) in a PCD (FCMD) while core denotes the radiative core (if any).  

Like \citetalias{SYT}, we assume that the convective envelope rotates as a solid body and parametrize the differential rotation with an average across the envelope as
$\Delta \Omega_\mathrm{env} \approx R_\mathrm{env}/R_\ast\,\Omega_\mathrm{env}$ \footnote{\textcolor{black}{We note that $R_\mathrm{{env}}$ is the width of the envelope (and not the location of the base of the envelope). The location of the base of the convective envelope is given by $R_\ast - R_\mathrm{env} = R_\mathrm{core}$}.}. This definition ensures that $\Delta \Omega_\mathrm{env} \xrightarrow{} \Omega_\mathrm{env}$ for FCMDs and that the smaller the thickness of the envelope, the lower its average differential rotation. This is the $\Omega$ term in our $\alpha-\Omega$ dynamo, which converts poloidal fields into toriodal fields. The dynamo is completed when the regeneration term or the $\alpha$ term in the $\alpha-\Omega$ dynamo converts toroidal fields into poloidal fields. The equilibrium equations remain the same as used by \citetalias{SYT}:
\begin{center}
\begin{equation}
\label{dyn1}
\frac{\mathrm{d}B_\mathrm{\phi}}{\mathrm{d}t} =\Delta\Omega_\mathrm{env} B_\mathrm{p} - \frac{B_\mathrm{\phi}}{\tau_\mathrm{\phi}} = 0
\end{equation}
\end{center}
and
\begin{center}
\begin{equation}
\label{eq:dyn3}
\frac{\mathrm{d}B_\mathrm{p}}{\mathrm{d}t} =\frac{\Gamma}{R_\ast} B_\mathrm{\mathrm{\phi}} - \frac{B_\mathrm{p}}{\tau_\mathrm{p}} = 0,
 \end{equation}
\end{center}
where $B_\mathrm{\phi}$ is an average toroidal component of the magnetic field in the star, $B_\mathrm{p}$ is the poloidal component of the magnetic field,  $\tau_\mathrm{\phi}$ and $\tau_\mathrm{p}$ are the time-scales on which the poloidal and toroidal magnetic field components are destroyed and $\Gamma$ is the dynamo regeneration term. We define $\Gamma$ similarly to \cite{1955ApJ...122..293P} and \cite{1981ARA&A..19..115C} so that 
\begin{center}
\begin{equation}
\label{eq:Gamma}
\Gamma \approx \tau_\mathrm{t}v_\mathrm{t}\omega_\mathrm{t} = \gamma v_\mathrm{c},
 \end{equation}
\end{center}
where $\tau_\mathrm{t}$ is the turbulent turnover time-scale, $v_\mathrm{t} \approx v_\mathrm{c}$ is the velocity of turbulent cells, $v_\mathrm{c}$ is the convective velocity and $\omega_\mathrm{t}\propto\Omega$ is the vorticity of the eddies parallel to $v_\mathrm{c}$. We write $\gamma$ similarly to equation~(5) of \citetalias{SYT} (see also their section~2.1)
\begin{center}
\begin{equation}
\label{eq:gamma}
\gamma = f_1 \frac{M_\ast}{0.35 M_\odot} \sqrt{\frac{R_\ast}{g}}\: \Omega_\mathrm{env},
 \end{equation}
\end{center}
where $f_1$ is a parameter of the order unity which we calibrate in Section~\ref{subs:calibration}. Here ${M_\ast}/{0.35 M_\odot}$ is the mass that undergoes cyclonic turbulence, normalized to $0.35 M_\odot$ which we take to be the maximum mass of a FCMD. We assume that turbulent instabilities turn over in the shortest time-scale possible. We take this to be the sound-crossing time over the stellar radius \citep{2002A&A...381..923S} and define $\tau_\mathrm{t}\approx\sqrt{{R_\ast}/{g}}$, where $g$ is the surface gravity.

\subsection{Wind mass loss calculation}
We extend the arguments of \citetalias{SYT} and assume that the energy input by differential rotation in the envelope is comparable to its rotational kinetic energy, but make careful modifications to take into account the fact that the star is only partly convective. The rate at which energy is fed into the shear in the convective zone is given by
\begin{center}
\begin{equation}
\label{eq:Lw}
L_\mathrm{w} \approx\frac{ \frac{1}{2} I_\mathrm{env}\left(\frac{R_\mathrm{env}}{R_\ast}\Omega_\mathrm{env}\right)^2} {\tau_\mathrm{\nu}}.
 \end{equation}
\end{center} 
where $I_\mathrm{env}$ is the moment of inertia of the envelope and $\tau_\nu$ is the viscous-time scale given by
\begin{center}
\begin{equation}
\label{eq:taunu}
\tau_\mathrm{\nu} = \frac{R_\mathrm{env}^2}{\nu},
 \end{equation}
\end{center} 
where $\nu$ is the convective viscosity. \citetalias{SYT}, following \cite{Zangrilli1997}, assumed that an arbitrary 10 percent of the deposited energy escapes as winds, and the rest is radiated away. We also implement this factor here. However, the fraction of energy escaping in winds should change when a star transits from a fully convective PMS star to a main-sequence (MS) star. Similarly, this fraction should be different for stars of different masses. These effects are encapsulated by the coronal temperature $T_\mathrm{cor}$. We argue that a larger fraction of the total shear luminosity is radiated away for a larger $T_\mathrm{cor}$. \cite{2015A&A...578A.129J} have used X-ray emission data of low-mass stars to derive how $T_\mathrm{cor}$ scales with stellar mass and spin. They find that for MS stars $T_\mathrm{cor}\propto M_\ast^{0.6}$  in the saturated regime and $T_\mathrm{cor}\propto M_\ast^{-0.42}\Omega_\ast^{0.52}$ in the unsaturated regime. However, these relations do not encapsulate the behaviour of $T_\mathrm{cor}$ of contracting PMS stars. This is extremely difficult to estimate reliably so we introduce an additional reduction of $(M_\mathrm{env}/M_\ast)^2 (M_\ast/M_\odot)^{-0.25}$ in the wind mass loss rate for partly convective stars. The term $(M_\mathrm{env}/M_\ast)^2$ illustrates that the fraction of wind energy reduces drastically when the star transits from its fully convective PMS phase to the partly convective MS phase. The term $(M_\ast/M_\odot)^{-0.25}$ illustrates that a larger fraction of the energy is radiated away in more massive stars. The exponents are ad hoc but yield realistic wind mass-loss rates for the Sun at its current age $t_\odot$. We write
\begin{center}
\begin{equation}
\label{eq:lw2mdot}
0.1\left(\frac{M_\mathrm{env}}{M_\ast}\right)^2\left(\frac{M_\ast}{M_\odot}\right)^{-0.25}L_\mathrm{w} \approx \frac{GM_\ast\Dot{M}_w}{R_\ast},
 \end{equation}
\end{center} 
where $G$ is Newton's gravitational constant. 
We then obtain an expression for the wind mass-loss rate 
\begin{center}
\begin{equation}
\label{eq:mlconv0}
\Dot{M}_\mathrm{w} =  f_\mathrm{corot}f_\mathrm{DZ} \left(\frac{M_\mathrm{env}}{M_\ast}\right)^2\left(\frac{M_\ast}{M_\odot}\right)^{-0.25} \frac{I_\mathrm{env}}{20} \frac{\Omega_\mathrm{env}^2 \nu}{G R_\ast M_\ast},
\end{equation}
\end{center}
where $f_\mathrm{DZ}$ is a dimensionless quantity, less than 1, which measures the effect of a dead zone on the star's mass-loss rate. The calculation of dead zones stays identical to that for fully convective stars and has been explained in detail by \citetalias{SYT}. The parameter $f_\mathrm{corot}$ is a dimensionless quantity that measures an enhancement of the winds (as well as a reduction of the Alfvén radius, described in Section~\ref{sec:aml}). The expression for $\nu$ is the same as that of \citetalias{SYT}:
\begin{center}
\begin{equation}
\label{eq:nu}
\nu \approx \frac{1}{3} v_\mathrm{c}l_\mathrm{c} \:\mathrm{min}\left(\left(\frac{2\pi f_2}{\tau_\mathrm{c}\Omega_\mathrm{env}}\right)^{p},\:1\right),
\end{equation}
\end{center}
where 
\begin{center}
\begin{equation}
\label{eq:vc}
v_\mathrm{c} \approx \left(\frac{L_\ast R_\ast}{\eta M_\ast}\right)^{1/3}.
 \end{equation}
\end{center}
Here $L_\ast$ is the luminosity of the star, $\eta \approx 3R_\mathrm{env}/l_\mathrm{c} \approx 30$ is a constant \citep{Campbell1983} and $l_\mathrm{c}$ is the mixing length. The factor $\mathrm{min}(({2\pi f_2}/{\tau_\mathrm{c}\Omega_\mathrm{env}})^{p},\:1)$ quantifies the curtailment in the convective viscosity for a rapidly rotating star (see section~2.2 of \citetalias{SYT} for details). The factor $f_2\gtrsim 1$ is a free parameter that dictates the $\Omega_\mathrm{env}$ at which $\nu$ is curtailed and $p\geq0$ measures how strongly the convective viscosity is curtailed when $\Omega_\mathrm{env}\tau_\mathrm{c} > 2\pi f_2$ ($p$ is calibrated in Section~\ref{subs:calibration}). Now $\Dot{M}_\mathrm{w}$ becomes
\begin{center}
\begin{equation}
\label{eq:mlconv}
\Dot{M}_\mathrm{w} = f_\mathrm{corot}f_\mathrm{DZ} \left(\frac{M_\mathrm{env}}{M_\ast}\right)^2\left(\frac{M_\ast}{M_\odot}\right)^{-0.25} \frac{I_\mathrm{env}}{60} \frac{\Omega_\mathrm{env}^2}{G R_\ast M_\ast}  v_\mathrm{c}l_\mathrm{c} \:\mathrm{min}\left(\left(\frac{2\pi f_2}{\tau_\mathrm{c}\Omega_\mathrm{env}}\right)^{p},\:1\right).
\end{equation}
\end{center}

\subsection{Alfvén radius, magnetic field, and angular momentum loss rate }
\label{sec:aml}
The winds that emerge from the star carry off angular momentum as if forced to corotate with it to the Alfvén radius, where  the kinetic energy density in the wind equals the magnetic energy density and so
\begin{center}
\begin{equation}
\label{eq:vA}
v^2_\mathrm{A} \approx \frac{B_\mathrm{p}(R_\mathrm{A})^2}{4\pi\rho_\mathrm{A}},
\end{equation}
\end{center}
where the subscript A denotes the Alfvén surface and $B_\mathrm{p}$ is the dipole component of the magnetic field. \citetalias{SYT} assumed that the wind density $\rho_\mathrm{A}$ at the Alfvén surface is spherically symmetric so that 
\begin{center}
\begin{equation}
\label{eq:rhoA1}
\rho_\mathrm{A} = \frac{\Dot{M}_\mathrm{w}}{4\pi R_\mathrm{A}^2 v_\mathrm{A}}.
\end{equation}
\end{center}
However, because of our implementation of the dead zones, not only does $\Dot{M}_\mathrm{w}$ reduce, but there is also a non-spherical wind mass loss profile (see \citealt{Garraffo2015} who show that magnetized winds are not distributed evenly at the stellar surface). This leads to a decrease in the surface area from which winds escape and consequently increases $\rho_\mathrm{A}$. We measure this decrease in surface area by redefining $\rho_\mathrm{A}$ as
\begin{center}
\begin{equation}
\label{eq:rhoA2}
\rho_\mathrm{A} = \frac{\Dot{M}_\mathrm{w}}{f_\mathrm{non-sph}4\pi R_\mathrm{A}^2 v_\mathrm{A}},
\end{equation}
\end{center}
where $f_\mathrm{non-sph}$ can be calculated if we know $f_\mathrm{DZ}$. \cite{1987MNRAS.226...57M} defined $\mathrm{sin}^2\theta\equiv f_\mathrm{DZ}=R_\ast/r_\mathrm{DZ}$, where $r_\mathrm{DZ}$ is the equatorial radius of the dead zone. \textcolor{black}{A larger $r_\mathrm{DZ}$ leads to a larger dead zone and, as a consequence, a smaller $f_\mathrm{DZ}$}. So the fractional surface of the wind-emitting region of the star can be calculated as
\begin{center}
\begin{equation}
\label{eq:fnonsph}
f_\mathrm{non-sph} = \epsilon \int_{0}^{\mathrm{sin}^{-1}(\sqrt{f_\mathrm{DZ}})} \mathrm{sin}\theta\, \mathrm{d}\theta = \epsilon\left(1 - \sqrt{\frac{r_\mathrm{DZ} - R_\ast}{R_\ast}}\right),
\end{equation}
\end{center}
where $\epsilon\ll 1$. 
However, the simulation results of \cite{Garraffo2015} show that winds are emitted only at certain latitudes where there are star-spots. This leads to a further mass-, radius- and likely spin-dependent reduction in the fraction of contributing surface area. This is extremely difficult to model so we simply incorporate it into $\epsilon$ by  
\begin{center}
\begin{equation}
\label{eq:epsilon}
\epsilon = 0.04\left(\frac{M_\odot}{M_\ast}\right)^\delta.
\end{equation}
\end{center}
Note that we introduce a new parameter $\delta$ that governs $\epsilon$. \textcolor{black}{This equation illustrates that, with increasing stellar mass, $\epsilon$ decreases and so the wind-emitting region on the stellar surface deviates further from spherical symmety. We note that starspot coverage and other activity proxies are thought to also decline as a function of age \citep{1972ApJ...171..565S} and stellar Rossby number \citep{2018MNRAS.479.2351W}. In equation~(\ref{eq:epsilon}) we present a simple case where the spherically symmetric wind coverage scales only with the star’s mass which, at first glance, may not appear to be physical. Although stellar activity evolves with age and Rossby number and as a result, equation~(\ref{eq:epsilon}) should have such dependencies incorporated, these are absorbed in the expression for torque (equation~\ref{eq:jdot_f2_epsil2_PMS}) which has an age- and Rossby number dependence because our main goal in this paper is to carefully model the spindown torque and not activity.}
 We now assume that the magnetic field falls off as a dipole everywhere (although see \citealt{Garraffo2015, Garraffo2016} for a discussion on multipolar contribution to the magnetic field) so that 
\begin{center}
\begin{equation}
\label{eq:bpR}
B_\mathrm{p}(R) = B_\mathrm{p}(R_\ast)\left(\frac{R_\ast}{R}\right)^3.
\end{equation}
\end{center}
Unlike \citetalias{SYT} we define the velocity at the Alfvén surface to be the escape velocity at the Alfvén radius $R_\mathrm{A}$ such that
\begin{center}
\begin{equation}
\label{eq:vA2}
v_\mathrm{A} = \sqrt{\frac{2GM_\ast}{R_\mathrm{A}}}.
\end{equation}
\end{center}
Because of this slight change in defining $v_\mathrm{A}$,  $R_\mathrm{A}$ differs from equation~(18) of \citetalias{SYT}:
\begin{center}
\begin{equation}
\label{eq:raold}
R_\mathrm{A} = f_\mathrm{non-sph}^{-2/7}\frac{R_\ast^{12/7}( B_\mathrm{p}(R_\ast))^{4/7}}{\Dot{M}_\mathrm{w}^{2/7} (2GM_\ast)^{1/7}}.
\end{equation}
\end{center}
The calculations of TP and \cite{Zangrilli1997} were for stars whose corotation radius $R_\mathrm{\Omega}\equiv (GM_\ast/\Omega^2)^{1/3}$ was larger than the Alfvén radius, so that the winds emitted from the photosphere were constantly decelerated. The calculations of \citetalias{SYT} did not take into account the scenario where $R_\mathrm{\Omega}<R_\mathrm{A}$ which is observed in many fast-spinning stars \citep{2018MNRAS.475L..25V}. \cite{1995MNRAS.273..146R} considered the case where $R_\mathrm{\Omega}<R_\mathrm{A}$ in which case winds are accelerated centrifugally beyond $R_\mathrm{\Omega}$ and can have velocities greater than the escape velocity at the Alfvén radius. This would lead to an enhancement in the wind mass loss and a reduction in $R_\mathrm{A}$. We model the effect as follows. 
When $R_\mathrm{A}>R_\mathrm{\Omega}$, owing to centrifugal acceleration the kinetic energy of the winds increases. This leads to an enhancement in the winds which we model with $\Dot{M}_\mathrm{w}\propto v^{2} \propto (R_\mathrm{A}/R_\mathrm{\Omega})$. This is taken into account in the factor $f_\mathrm{corot}$ in equation~(\ref{eq:mlconv}). We assume that the wind velocity at the Alfvén surface increases leading to a decrease in $R_\mathrm{A}$ such that it cancels the increase in $\Dot{M}_\mathrm{w}$ and $\Dot{J}_\mathrm{w}\propto \Dot{M}_\mathrm{w}R_\mathrm{A}^2$ remains unchanged. \textcolor{black}{This is done solely for convenience so that we do not have to deal with an additional factor that affects $\dot{J}_\mathrm{w}$. However, in reality, $\Dot{M}_\mathrm{w}$ may be significantly altered for fast rotators.} To ensure this, we add a factor $f_\mathrm{corot}^{-3/14}$ to equation~(\ref{eq:raold}) which becomes
\begin{center}
\begin{equation}
\label{eq:ra}
R_\mathrm{A} = f_\mathrm{corot}^{-3/14}f_\mathrm{non-sph}^{-2/7}\frac{R_\ast^{12/7}( B_\mathrm{p}(R_\ast))^{4/7}}{\Dot{M}_\mathrm{w}^{2/7} (2GM_\ast)^{1/7}},
\end{equation}
\end{center}
where 
\begin{center}
\begin{equation}
\label{eq:fcorot}
f_\mathrm{corot} = \mathrm{max}\left( \left(\frac{R_\mathrm{A}}{R_\mathrm{\Omega}}\right),\;1 \right).
\end{equation}
\end{center}
The dynamo mechanism leads to a surface poloidal field which is given by equation (4.10) of TP modified for a partly convective star:
\begin{center}
\begin{equation}
\label{eq:bpconv}
B_\mathrm{p}(R_\ast) = 10\gamma v_\mathrm{c}\sqrt{4\pi \rho_\mathrm{env}},
\end{equation}
\end{center}
where $\rho_\mathrm{env} = M_\mathrm{env}/(4/3\pi R_\ast^3 - 4/3\pi R_\mathrm{core}^3)$ is the mean density of the convective envelope.   Beyond the Alfvén radius, the winds escape freely, carrying away angular momentum. So we write the angular momentum loss as
\begin{center}
\begin{equation}
\label{eq:jdot}
\Dot{J}_\mathrm{w} = - \Dot{M}_\mathrm{w} R_\mathrm{A}^2 \Omega_\mathrm{env}.
\end{equation}
\end{center}

\begin{figure}

\includegraphics[width=0.45\textwidth]{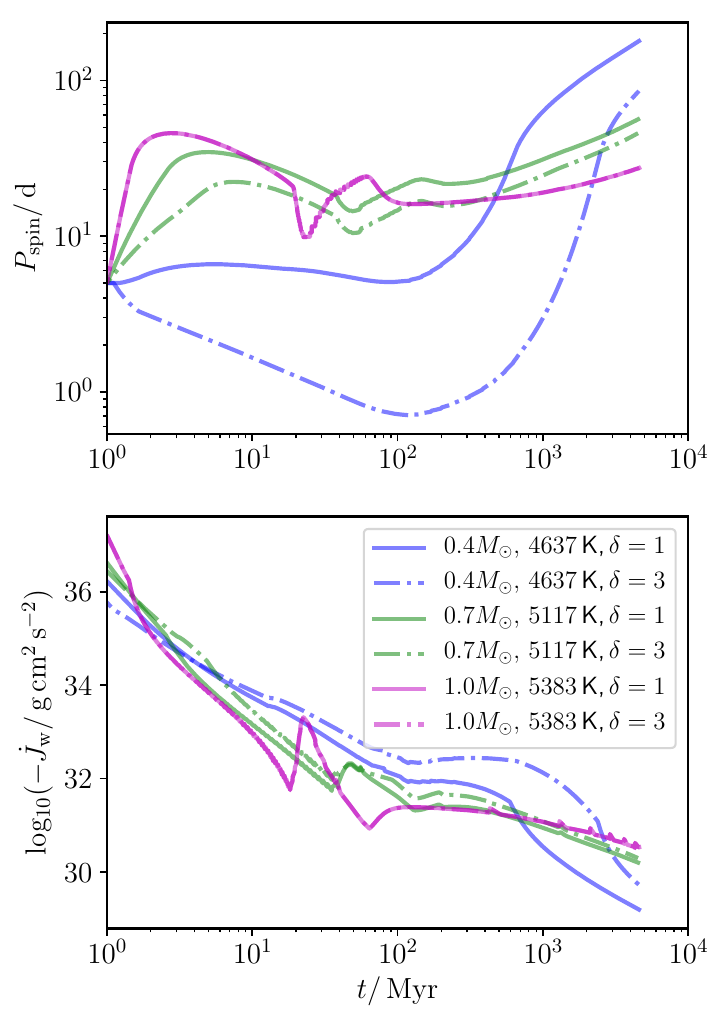}
\caption{The evolution of the  envelope spin $P_\mathrm{spin}$ and MB torque $\dot{J}_\mathrm{w}$ with time for $0.4M_\odot$, $0.7M_\odot$ and $1.0M_\odot$ stars with $P_\mathrm{spin,\,initial}=5\,$d and $\tau_\mathrm{dl}=1\,$Myr for different $\delta$. The free parameters have been set as $f_2=2$ and $f_\mathrm{shear}=0.3$ (Section~\ref{subs:calibration}). The effective temperatures at $\tau_\mathrm{dl}$ are shown in the legend. The two tracks for the $1.0M_\odot$ star coincide because of the definition of $\delta$ in equation~(\ref{eq:epsilon}). }
\label{fig:PMS_analysis}
\end{figure}

\textcolor{black}{A point of discussion may be the reliability of the ad hoc factor $(M_\mathrm{env}/M_\ast)^2(M_\ast/M_\odot)^{-0.25}$ in equations~(\ref{eq:lw2mdot} to \ref{eq:mlconv}). This is because, for stars with larger convective envelopes (low-mass MS and PMS stars), this induces a significant enhancement in winds at early times, which can significantly affect subsequent results and the interpretation of the relative importance of competing internal effects. The torque $\dot{J}_\mathrm{w}$ can be written as\begin{center}
\begin{equation}
\label{eq:jdot_f2_epsil2_PMS}
 -\Dot{J}_\mathrm{w} \propto     \begin{cases}
        f_2^2 \left(\frac{M_\mathrm{env}}{M_\ast}\right)^{\frac{6}{7}}\left(\frac{M_\ast}{M_\odot}\right)^{\frac{16\delta-3}{28}}, & \text{if   }\; ({2\pi f_2}/{\tau_\mathrm{c}\Omega_\mathrm{env}})\leq 1 \\
        \\
\left(\frac{M_\mathrm{env}}{M_\ast}\right)^{\frac{6}{7}}\left(\frac{M_\ast}{M_\odot}\right)^{\frac{16\delta-3}{28}}, & \text{otherwise}.
\end{cases}
\end{equation}
\end{center} This equation contains all the uncertain parameters our torque depends on. In Section~\ref{subs:calibration} we calibrate $f_1$ and set $f_2$ and $\delta$ as free parameters of order unity. For now, we ignore $f_2$ as it only affects $\dot{J}_\mathrm{w}$ by a factor of a few. We plot our estimates of $\dot{J}_\mathrm{w}$ for different values of $\delta$ in Fig.~\ref{fig:PMS_analysis} and compare those with the results of \cite{2023AJ....165..182K}. They find spin-down torques of about $10^{37}$~erg at about 1~Myr, $10^{36}$~erg at about 2~Myr and $10^{35}$~erg at about 10~Myr, with initial spin-down torques increasing with increasing stellar mass ($T_\mathrm{eff}$, their fig.~7). Their estimates agree with ours in Fig.~\ref{fig:PMS_analysis}. It is seen that although different values of $\delta$ lead to different envelope spin evolution, the torques up until about 100~Myr are almost the same for a given $M_\ast$ (the behaviour of the models with respect to their structure is described in Section~\ref{subs:analysis} and Fig.~\ref{fig:analysis}). This illustrates that our factor $(M_\mathrm{env}/M_\ast)^2(M_\ast/M_\odot)^{-0.25}$ in $\dot{J}_\mathrm{w}$ produces results in general agreement with observations of torque at early times. We study the dependence of our tracks on $\delta$ in Section~\ref{subs:param_analysis}. }

\subsection{Core-envelope shear}
\label{subs:shear}
We model the shear at the core-envelope boundary based on the arguments of \citet[their section~4]{Zangrilli1997}. We assume for simplicity that the radiative core spins as a solid body with $\Omega_\mathrm{core}$. Thus the differential rotation at the core-envelope boundary is 
\begin{center}
\begin{equation}
\label{eq:diffrot}
\Delta \Omega_\mathrm{shear} = \Omega_\mathrm{core} - \Omega_\mathrm{env}.
 \end{equation}
\end{center}
When $\Delta\Omega_\mathrm{shear}\neq 0$, any convective cell that overshoots from the envelope to the core experiences the action of a toroidal field, which is a function of  $\Delta\Omega_\mathrm{shear}$, at the boundary. A consequence of this effect is shear that spins down the core if  $\Delta\Omega_\mathrm{shear}>0$ and vice versa. In other words, differential rotation at the core-envelope boundary creates an angular momentum transport mechanism between the two regions.

According to the arguments of \cite{Zangrilli1997}, the toroidal field generated at the boundary layer decays quickly beyond the thickness of the boundary layer $H_\mathrm{B}$ (the grey region in the right illustration Fig.~\ref{fig:cartoon}). This is because we assume that the differential rotation responsible for the strong toroidal field only persists in the boundary layer. So we define the decay time-scale for the toroidal field as
\begin{center}
\begin{equation}
\label{eq:tauphi}
\tau_\phi = \frac{10H_\mathrm{B}}{v_\phi},
 \end{equation}
\end{center}
where $v_\phi = B_\phi/\sqrt{4\pi\rho_\mathrm{B}}$. The factor of 10 arises because we assume that the field decays by magnetic buoyancy, for which we assume a timescale about 10 times longer than the Alfvén-wave crossing time (see section~4 of TP for details). However, the poloidal field is generated throughout the convective layer because of the $\alpha$-effect and only decays beyond $R_\mathrm{env}$. So the decay time-scale for the poloidal field is
\begin{center}
\begin{equation}
\label{eq:taup}
\tau_\mathrm{p} = \frac{10R_\mathrm{env}}{v_\mathrm{p}},
 \end{equation}
\end{center}
where $v_\mathrm{p} = B_\mathrm{p}/\sqrt{4\pi\rho_\mathrm{B}}$. Now, using equations~(\ref{dyn1}) and (\ref{eq:dyn3}), we get
\begin{center}
\begin{equation}
\label{eq:vp}
v_\mathrm{p} = 10\gamma^{2/3} v_\mathrm{c}^{2/3}\left(\frac{R_\mathrm{env}}{R_\ast}\right)^{2/3} \Delta \Omega_\mathrm{shear}^{1/3}H_\mathrm{B}^{1/3}
 \end{equation}
\end{center}
and
\begin{center}
\begin{equation}
\label{eq:vphi}
v_\phi = 10\gamma^{1/3} v_\mathrm{c}^{1/3}\left(\frac{R_\mathrm{env}}{R_\ast}\right)^{1/3} \Delta \Omega_\mathrm{shear}^{2/3}H_\mathrm{B}^{2/3}.
 \end{equation}
\end{center}
The rate of angular momentum transfer between the core and the envelope is given by
\begin{center}
\begin{equation}
\label{eq:jdotshear}
\Dot{J}_\mathrm{shear} = \left(\frac{B_\mathrm{p}B_\phi}{4\pi}\right)V_\mathrm{B} = f_\mathrm{shear} 100\gamma v_\mathrm{c} \left(\frac{R_\mathrm{env}}{R_\ast}\right)\Delta\Omega_\mathrm{shear} H_\mathrm{B}^2 R_\mathrm{core}^2\:4\pi\rho_\mathrm{B},
 \end{equation}
\end{center}
where $f_\mathrm{shear}\leq1$ is a free parameter that measures the efficiency of shear to transport angular momentum, $V_\mathrm{B} = 4\pi H_\mathrm{B}R_\mathrm{core}^2$ is the volume of the boundary layer and $\rho_\mathrm{B}= M_\mathrm{B}/V_\mathrm{B}$, where $M_\mathrm{B}$ is the mass of the boundary layer (also see equation~13 of \citealt{Wickramasinghe2013}). We define a simple empirical expression of $H_\mathrm{B}$ as
\begin{center}
\begin{equation}
\label{eq:hb}
H_\mathrm{B} \equiv 0.0645R_\mathrm{env}. 
\end{equation}
\end{center}
The factor 0.0645 gives $H_\mathrm{B}\approx0.02R_\odot$ at $t\approx4.6$ Gyr for a $1M_\odot$ star. This is close to the size of the solar tachocline as estimated by \cite{2003ApJ...585..553B}. With equation~(\ref{eq:hb}) we can estimate the thickness of the boundary layer (tachocline) of a star of a given mass at all evolutionary times.

\subsection{Calibration of free parameters and a summary of updates}
\label{subs:calibration}

The parameters $f_1$, $f_2$, $p$, $\delta$, and $f_\mathrm{shear}$ must be estimated before we can use our model to evolve PCD spins. We keep $f_2$, $\delta$ and $f_\mathrm{shear}$ as free parameters and fix $f_1$ and $p$ as follows.
\begin{itemize}
    \item $f_1$: Using equation~(\ref{eq:gamma}) in equation~(\ref{eq:bpconv}), we obtain $f_1$ as a function of $t,\,B_\mathrm{p}, M_\ast, R_\ast, L_\ast$ and $P_\mathrm{spin}$. Using solar data from Table~\ref{tab:solar} and $M_\mathrm{env}$ and $R_\mathrm{env}$ at $t_\odot$ using STARS, we obtain $f_1\approx 0.115$.

    \item $p$: This parameter measures how strongly convective viscosity is reduced for a fast-spinning star (when $\Omega_\mathrm{env}\tau_\mathrm{c}>2\pi f_2$). We fixed $p=4$ in \citetalias{SYT} (see their section~3.1) which gave $\Dot{J}_\mathrm{w} \propto \Omega$ in the saturated regime and $\Dot{J}_\mathrm{w} \propto \Omega^3$ in the unsaturated regime. Owing to the redefinition of $R_\mathrm{A}$ in equation~(\ref{eq:ra}), we now have

\begin{center}
\begin{equation}
\label{eq:jdotprop_cases}
\Dot{J} \propto
    \begin{cases}
        \Omega_\mathrm{env}^{3-\frac{3p}{7}}, & \text{when   }\; 2\pi f_2/\tau_\mathrm{c}\Omega_\mathrm{env} \; \leq \; 1\; \mathrm{(saturated)},\\
        \\
        \Omega_\mathrm{env}^{3}, & \text{otherwise}\; \mathrm{(unsaturated)}.
    \end{cases}
\end{equation}
\end{center}
which gives us the desired proportionality when $p=14/3$.
\end{itemize}
Our three free parameters are tabulated in Table~\ref{tab:table_param} and we summarise the changes implemented in our MB modelling since the work of \citetalias{SYT} in Table~\ref{tab:table_change}.

\begin{table}
\caption{Solar data used in this work.}
\label{tab:solar}
\begin{tabular}{lc}
\hline
Quantity & Value \\
\hline
$P_\odot$& $26.09\,$d\\
$t_\odot$   &  4.6 Gyr\\
$\Dot{M}_\mathrm{w\odot}$   &  $2\times10^{-14}\,M_\odot\,\mathrm{yr}^{-1}$\\
$B_\mathrm{p\odot}$   &  $1$ G\\
$R_\mathrm{A\odot}$   &  $10\,\mathrm{to}\,20\,R_\odot$\\
\hline
\end{tabular}
\end{table}

\begin{table}
	
	\caption{A table of parameters and a description of their estimates.}
	\label{tab:table_param}
	\begin{tabular}{ll} 
		\hline
		 Parameter & Description  \\
		\hline
 
          $f_2$, equation~(\ref{eq:nu})& Critical Rossby number\\

           \\

          $\delta$, equation~(\ref{eq:epsilon})& Probes the behaviour of the wind-emitting surface\\

          \\
          $f_\mathrm{shear}$, equation~(\ref{eq:jdotshear})& Efficiency of shear\\
		\hline
	\end{tabular}
\end{table}

\begin{table*}
	\centering
	\caption{A summary of the differences in the magnetic braking implementation of \protect\citetalias{SYT} and this work with the calibrated values of certain parameters.}
	\label{tab:table_change}
	\begin{tabular}{lccr} 
		\hline
		 Parameter & In \protect\citetalias{SYT} & In this work  & Description of change/calibration\\
		\hline
         Convective length $l_\mathrm{c}$ & $\displaystyle\frac{R_\ast}{10}$ & $\displaystyle\frac{R_\mathrm{env}}{10}$ & Reduction for smaller convective envelopes\\
\\  
         Shear in convective zone $\Delta\Omega_\mathrm{env}$& $\Omega\equiv\Omega_\mathrm{env}$ & $\displaystyle\frac{R_\mathrm{env}}{R_\ast}\Omega_\mathrm{env}$ & Reduction in shear for smaller convective envelopes\\
  \\
		 Wind mass loss $\Dot{M}_\mathrm{w}$ & $\displaystyle{f_\mathrm{DZ} \frac{1}{600} \frac{R_\ast}{G} \Omega^2 v_\mathrm{c}l_\mathrm{c} \:\mathrm{min}\left(\left(\frac{2\pi f_2}{\tau_\mathrm{c}\Omega}\right)^{p},\:1\right)}$ & Equation~(\ref{eq:mlconv})& Reduction in $\tau_\nu$, parametrization of wind escape fraction \\
  \\
		 Measure of viscosity curtailment $p$ & 4 & $\displaystyle\frac{14}{3}$ & Due to the redefinition of $R_\mathrm{A}$ \\   

   \\
   $f_1$, equation~(\ref{eq:gamma})& Free parameter & 0.115& Calibrated using $B_\mathrm{p\odot}$ at $t_\odot$\\  
\\
        Alfvén wind density $\rho_\mathrm{A}$ & $ \displaystyle{\frac{\Dot{M}_\mathrm{w}}{4\pi R_\mathrm{A}^2 v_\mathrm{A}}}$ & Equation~(\ref{eq:rhoA2})& Reduction in the surface area of wind emission\\
\\
        Wind velocity at Alfvén surface $v_\mathrm{A}$ & $\displaystyle{ \sqrt{{2GM_\ast}/{R_\ast}}}$ & Equation~(\ref{eq:vA2})& Escape velocity at Alfvén radius\\
\\
        Surface poloidal field $B_\mathrm{p}(R_\ast)$ & $\displaystyle{ 10\gamma v_\mathrm{c}\sqrt{4\pi \rho_\ast}}$ & Equation~(\ref{eq:bpconv})& Mean density of the convective envelope\\
\\        
        Alfvén radius $R_\mathrm{A}$& $\displaystyle{R_\ast\left(\frac{B_\mathrm{p}(R_\ast)^2R_\ast^2}{\Dot{M}_\mathrm{w}\sqrt{{2GM_\ast}/{R_\ast}}}\right)^{1/4}}$ & Equation~(\ref{eq:ra})& A different definition of $v_\mathrm{A}$\\        
		\hline
	\end{tabular}
\end{table*}

\section{Modelling of the spin evolution}
\label{sec:spind}
We can now model the spin evolution of any fully convective or partly convective star with a convective envelope and a radiative core if we know the time evolution of $L_\ast$, $R_\ast$, $M_\ast$, $I_\mathrm{core}$, $I_\mathrm{env}$, $M_\mathrm{core}$, $M_\mathrm{env}$, $R_\mathrm{core}$, $R_\mathrm{env}$ and $M_\mathrm{B}$. We use the Cambridge stellar evolution code \textsc{STARS} \citep{1973MNRAS.163..279E, 1995MNRAS.274..964P} to obtain these parameters as a function of time at solar metallicity. We assume, similarly to \citetalias{SYT}, that the winds do not reduce the stellar mass significantly and so keep $M_\ast$ constant with time. \textcolor{black}{We also assume that there is no radius inflation induced by the stellar magnetic field in our models (\citealt{2024ApJ...963...43M} and the references therein).} In other words, we assume that the spin evolution of the star does not alter any of the parameters above and these parameters only change because of non-rotational stellar evolution.

The equations that govern the angular momentum evolution of the core $J_\mathrm{core}$ and the envelope $J_\mathrm{env}$ are two ordinary differential equations:
\begin{center}
\begin{equation}
\label{eq:jcdot}
 \Dot{J}_\mathrm{core} =     \begin{cases}
        - \Dot{J}_\mathrm{shear} + \frac{2}{3}\Dot{M}_\mathrm{core} R_\mathrm{core}^2 \Omega_\mathrm{env}, & \text{if  }\; \Dot{M}_\mathrm{core}>0\\
        \\
- \Dot{J}_\mathrm{shear} + \frac{2}{3}\Dot{M}_\mathrm{core} R_\mathrm{core}^2 \Omega_\mathrm{core}, & \text{if  }\; \Dot{M}_\mathrm{core}\leq0
    \end{cases}
\end{equation}
\end{center}
and
\begin{center}
\begin{equation}
\label{eq:jedot}
 \Dot{J}_\mathrm{env} = \begin{cases}\Dot{J}_\mathrm{w} +  \Dot{J}_\mathrm{shear} - \frac{2}{3}\Dot{M}_\mathrm{core}R_\mathrm{core}^2 \Omega_\mathrm{env}, & \text{if  }\; \Dot{M}_\mathrm{core}>0\\
 \\
\Dot{J}_\mathrm{w} +  \Dot{J}_\mathrm{shear} - \frac{2}{3}\Dot{M}_\mathrm{core}R_\mathrm{core}^2 \Omega_\mathrm{core}, & \text{if  }\; \Dot{M}_\mathrm{core}\leq0
    \end{cases}
\end{equation}
\end{center}
Equation~(\ref{eq:jcdot}) illustrates that the angular momentum of the core is governed by shear (the first term) as well as the structural change of the star (the second term) when regions of the star change from being convective to being radiative, for instance, at the end of the PMS phase in F, G, K and (certain) M stars. We assume that there is no discontinuity in the $\Omega$ profile of the star at the tachocline so that the core boundary always spins with $\Omega_\mathrm{env}$. Equation~(\ref{eq:jcdot}) illustrates that the same phenomena plus magnetic braking govern the angular momentum evolution of the envelope. Now, under the assumption of solid-body rotation in the core and the envelope such that they rotate with an average $\Omega_\mathrm{core}$ and $\Omega_\mathrm{env}$ throughout, we obtain
\begin{center}
\begin{equation}
\label{eq:omegacdot}
 \Dot{\Omega}_\mathrm{core} =     \begin{cases}
        \displaystyle{\frac{- \Dot{J}_\mathrm{shear} + \frac{2}{3}\Dot{M}_\mathrm{core} R_\mathrm{core}^2 \Omega_\mathrm{env} -\Dot{I}_\mathrm{core}\Omega_\mathrm{core}}{I_\mathrm{core}}}, & \text{if  }\; \Dot{M}_\mathrm{core}>0\\
        \\
\displaystyle{\frac{- \Dot{J}_\mathrm{shear} + \frac{2}{3}\Dot{M}_\mathrm{core} R_\mathrm{core}^2 \Omega_\mathrm{core}-\Dot{I}_\mathrm{core}\Omega_\mathrm{core}}{I_\mathrm{core}}}, & \text{if  }\; \Dot{M}_\mathrm{core}\leq0
    \end{cases}
\end{equation}
\end{center}
and
\begin{center}
\begin{equation}
\label{eq:omegaedot}
 \Dot{\Omega}_\mathrm{env} = \begin{cases}
 \displaystyle{\frac{\Dot{J}_\mathrm{w} +  \Dot{J}_\mathrm{shear} - \frac{2}{3}\Dot{M}_\mathrm{core}R_\mathrm{core}^2 \Omega_\mathrm{env} -\Dot{I}_\mathrm{env}\Omega_\mathrm{env}}{I_\mathrm{env}}}, & \text{if  }\; \Dot{M}_\mathrm{core}>0\\
 \\
\displaystyle{\frac{\Dot{J}_\mathrm{w} +  \Dot{J}_\mathrm{shear} - \frac{2}{3}\Dot{M}_\mathrm{core}R_\mathrm{core}^2 \Omega_\mathrm{core} -\Dot{I}_\mathrm{env}\Omega_\mathrm{env}}{I_\mathrm{env}}}, & \text{if  }\; \Dot{M}_\mathrm{core}\leq0.
    \end{cases}
\end{equation}
\end{center}
We note that the core can indeed have a radial dependence on $\Omega$ \citep{Btrisey2023, Eggenberger2019}. This is encapsulated in our simple treatment of spins as averages such that $\Omega_\mathrm{core}>\Omega_\mathrm{env}$ is equivalent to the possibility of a larger $\Omega$ at the deep interior of a star.

\subsection{Spin evolution of a $0.5M_\odot$ and a $1.0M_\odot$ star}
\label{subs:analysis}

\begin{figure*}
\centering
\includegraphics[width=0.85\textwidth]{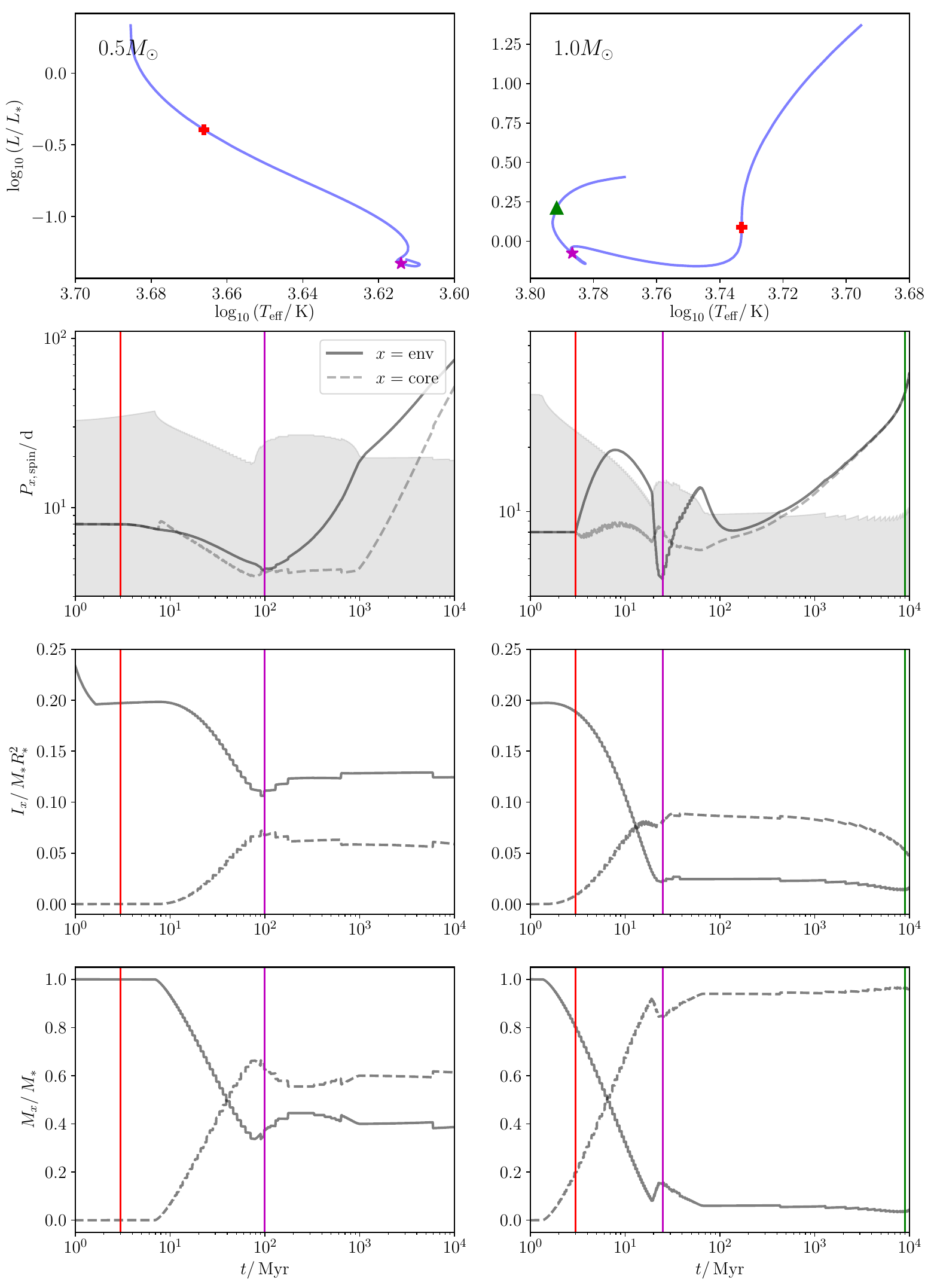}
\caption{The evolution of the core and the envelope spin $P_\mathrm{spin}$, moment of inertia $I$ and masses with time for a $0.5M_\odot$ and $1.0M_\odot$ star with $P_\mathrm{spin,\,initial}=8\,$d and $\tau_\mathrm{dl}=3\,$Myr. The free parameters have been set as $f_2=1.5$, $f_\mathrm{shear}=0.1$ and $\delta=1$. The red pluses (and the vertical lines of the same colour) denote the end of the disc-locking time. The magenta stars denote the minimum $P_\mathrm{env,spin}$ and the green triangles the terminal age main sequence. The shaded region in the $P_\mathrm{spin}-t$ plot denotes the region where $P_\mathrm{spin}\leq P_\mathrm{crit}\equiv \tau_\mathrm{c}/f_2$. }
\label{fig:analysis}
\end{figure*}

\textcolor{black}{Here we describe how the spin evolution of our stars, particularly at early times, depends on the evolution of their internal structure. Although these effects are present in the subsequent sections, later we shall focus on the effect of uncertain free parameters on our models.} 

We analyse the spin evolution of the core and envelope of a $0.5M_\odot$ and a $1.0M_\odot$ star in Fig.~\ref{fig:analysis}. These represent two qualitatively different PCD stars: the $0.5M_\odot$ star always has $I_\mathrm{env}>I_\mathrm{core}$ while in the $1.0M_\odot$ star $I_\mathrm{core}>I_\mathrm{env}$ after the PMS phase. 

We assume that the stars do not undergo any spin evolution before the disc-locking time $\tau_\mathrm{dl}$ \citep{Rebull2004}. After $\tau_\mathrm{dl}$, the $0.5M_\odot$ star is still fully convective. Until about 10 Myr $I_\mathrm{core}\approx0$ and $I_\mathrm{env}$ stays fairly constant. Beyond this, the radiative core emerges so  $I_\mathrm{env}$ begins to decrease and the envelope loses angular momentum to the core (the last term in equations~\ref{eq:jcdot} and \ref{eq:jedot}). The core gains angular momentum from the envelope and $I_\mathrm{core}$ increases. At the same time, the shear term (equation~\ref{eq:jdotshear}) operates on the core to slow its spin while spinning up the envelope. Overall, the star spins up because of the contraction of the PMS star.  This remains the case until about the magenta line at 100 Myr, beyond which the core and the envelope maintain an almost constant $I$. From then onwards the spin behaviour of the core and the envelope can be understood as follows. When $I_\mathrm{core}$ and $I_\mathrm{env}$ are relatively constant, we can write 
\begin{center}
\begin{equation}
\label{eq:omegac_const}
\Dot{\Omega}_\mathrm{core} = \frac{ - \Dot{J}_\mathrm{shear}}{I_\mathrm{core}}
 \end{equation}
\end{center}
and
\begin{center}
\begin{equation}
\label{eq:omegae_const}
\Dot{\Omega}_\mathrm{env} = \frac{ \Dot{J}_\mathrm{w} + \Dot{J}_\mathrm{shear}}{I_\mathrm{env}}.
 \end{equation}
\end{center}
The spin-down of the core is slower than the spin-down of the envelope when $ - \Dot{\Omega}_\mathrm{core} < - \Dot{\Omega}_\mathrm{env}$  so when $\Dot{\Omega}_\mathrm{core} >  \Dot{\Omega}_\mathrm{env}  $. Note that $\Dot{J}_\mathrm{w}<0$ and for spin-down $\Dot{\Omega}$s < 0. So using equations~(\ref{eq:omegac_const} and \ref{eq:omegae_const}) we obtain
\begin{center}
\begin{equation}
\label{eq:Iineq}
\;\frac{-\Dot{J}_\mathrm{w}}{\Dot{J}_\mathrm{shear}} \begin{cases}  \displaystyle{> \frac{I_\mathrm{env}+I_\mathrm{core}}{I_\mathrm{core}}} \:\mathrm{then}\: \Dot{\Omega}_\mathrm{core} >  \Dot{\Omega}_\mathrm{env}\:\text{(core spins down slower)},   \\\\
  \displaystyle{\leq \frac{I_\mathrm{env}+I_\mathrm{core}}{I_\mathrm{core}}}  \:\mathrm{then}\: \Dot{\Omega}_\mathrm{core} \leq  \Dot{\Omega}_\mathrm{env}\:\text{(core spins down faster).} \\
\end{cases}
 \end{equation}
\end{center}
The LHS of this expression is always positive for a more rapidly spinning core. For the $0.5 M_\odot$ star, \textcolor{black}{after an initial spin up due to structural changes}, the core does not change in its spin from 100~Myr to about 1~Gyr. This is because $\Dot{J}_\mathrm{shear}$ is too weak to spin down the core and MB is strong so the first inequality in (\ref{eq:Iineq}) holds. On the other hand, the envelope spins down rapidly by MB. Beyond about 1~Gyr, $I_\mathrm{core}$ decreases slightly allowing $\Dot{J}_\mathrm{shear}$ to slow the core down more rapidly. The envelope's spin-down rate is reduced because now it is spun down by MB but spun up by shear (the second case in expression~\ref{eq:Iineq}). However, the core and the envelope do not achieve corotation within the Galactic Age of about 12~Gyr \citep{2019NatAs...3..932G}. This depends on the free parameters as illustrated in Section~\ref{subs:param_analysis}.

For the $1M_\odot$ star, the PMS phase has ended by $\tau_\mathrm{dl}$ but the star is still dominantly convective. Owing to its strong wind mass loss (equation~\ref{eq:mlconv}), the envelope experiences a very strong MB spin down which dominates over its contraction-driven spin up. Spin-up dominates from about 10~Myr till about 20~Myr when $I_\mathrm{env}$ is at its minimum. At the same time, $I_\mathrm{core}$ increases and the core also gains angular momentum from the envelope which keeps it spinning faster than the envelope. Interestingly, it is seen that from about 20 to 30~Myr, the core is spinning slower than the envelope. This is a stellar-structure-dependent effect and is due to the very rapid spin-up of the envelope when $I_\mathrm{env}$ is at its minimum. Beyond this, the envelope quickly spins down. From 50~Myr, the envelope and the core are slowly brought into near corotation by shear. The equality in the second case of expression~(\ref{eq:Iineq}) illustrates the behaviour of the star when the spin-down rates of the core and the envelope are the same.  For the $1M_\odot$ star, beyond 30~Myr $(I_\mathrm{env}+I_\mathrm{core})/I_\mathrm{core} \approx 1$ because of a small $I_\mathrm{env}$. So shear and MB have similar strengths but MB is always stronger. In other words, the spinning-down envelope drags the core along with it. Because $(I_\mathrm{env}+I_\mathrm{core})/I_\mathrm{core} > 1$, this ensures that a structurally constant star always spins down.

\subsection{Dependence on free parameters}
\label{subs:param_analysis}

We now analyse the dependence of the spin-evolution trajectories on the free parameters in our model, $f_\mathrm{shear}$, $f_2$ and $\delta$, which govern the strength of shear at the boundary layer, the transition of the MB torque from the saturated to the unsaturated regime and the fraction of the wind-emitting surface of the star (see section~3 of \citetalias{SYT} for a detailed discussion of $f_2$).

\subsubsection{Shear uncertainty}
\label{subss:shear}

\begin{figure}
\centering
\includegraphics[width=0.4\textwidth]{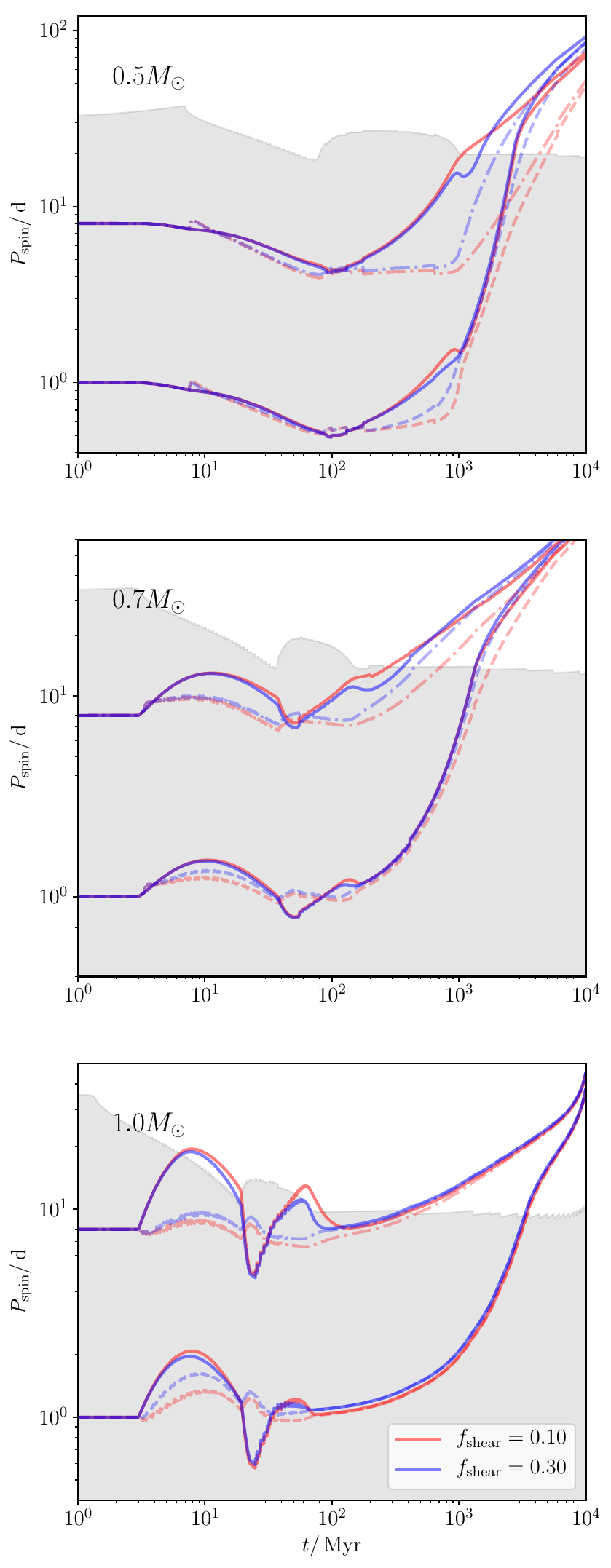}
\caption{The evolution of the core (dashed for the fast rotators, dash-dotted for the slow rotators) and the envelope (solid) spin $P_\mathrm{spin}$ for $0.5M_\odot$, $0.7M_\odot$ and $1.0M_\odot$ stars with $P_\mathrm{spin,\,initial}/\,\mathrm{d}\in\{1,\,8\}$ and $\tau_\mathrm{dl}=3\,$Myr for different $f_\mathrm{shear}$. The other free parameters have been set to $f_2=1.5$ and $\delta=1$. The shaded region in the respective plots denotes the region where $P_\mathrm{spin}\leq P_\mathrm{crit}\equiv \tau_\mathrm{c}/f_2$.}
\label{fig:shear_analysis}
\end{figure}
 We first keep the MB-affecting terms $f_2$ and $\delta$ fixed while varying $f_\mathrm{shear}$ (Fig.~\ref{fig:shear_analysis}). It is seen that a larger $f_\mathrm{shear}$ corresponds to a stronger angular momentum transfer mechanism between the core and the envelope, which leads to the core and the envelope attaining corotation slightly earlier in the $1.0M_\odot$ star. For the $0.5M_\odot$ star, a stronger shear leads to a lower degree of differential rotation between the core and the envelope after a Gyr, with $\Delta\Omega_\mathrm{shear}\approx0$ by 10 Gyr when $f_\mathrm{shear}=0.3$ for both initial $P_\mathrm{spin}$s. Changing the efficiency of shear leads to slight changes in the surface spin trajectory of only the slow-rotating $0.5M_\odot$ star, and only after a Gyr, while the $1.0M_\odot$ trajectories remain almost identical to each other. The effect is intermediate in the $0.7M_\odot$ star. In other words, our spin trajectories, particularly surface spins, are robust to changes in the efficiency of angular momentum exchange by shear. 

\subsubsection{Magnetic braking uncertainty}
\label{subss:mb}

\begin{figure}
\centering
\includegraphics[width=0.4\textwidth]{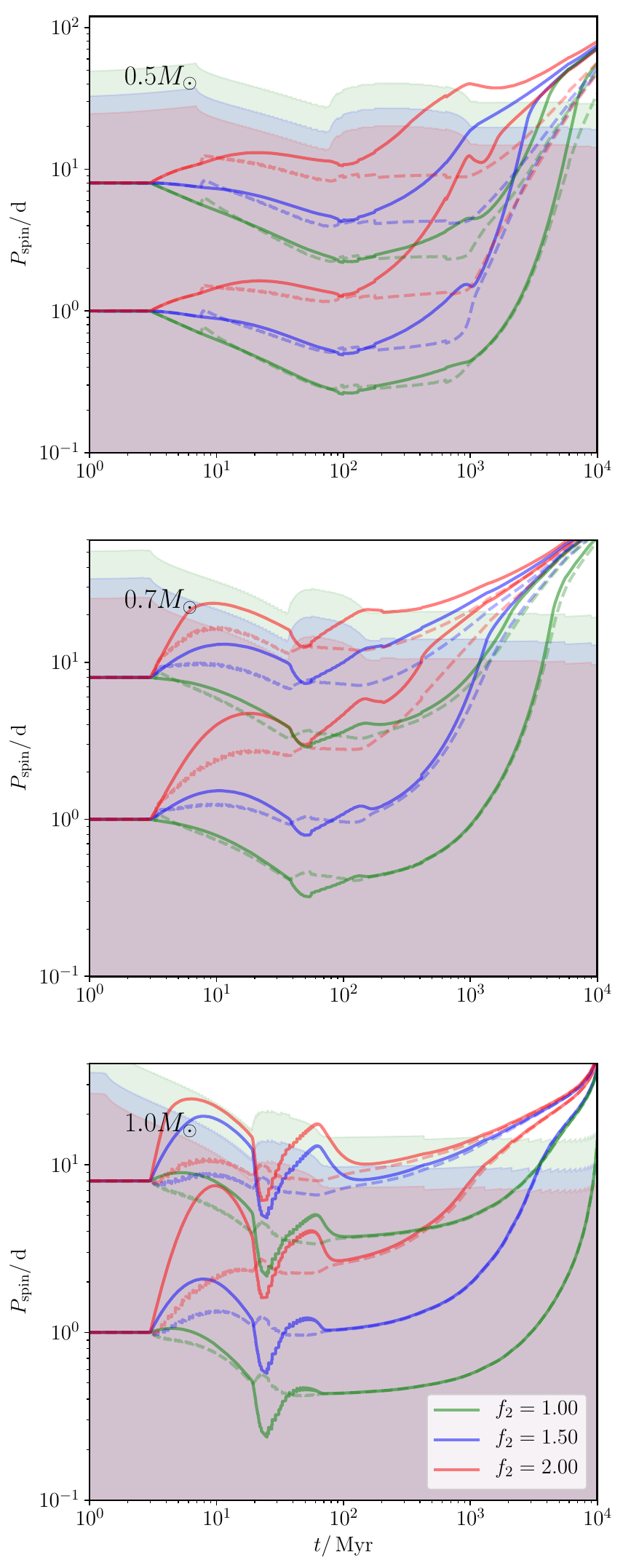}
\caption{The evolution of the core (dashed) and the envelope (solid) spin $P_\mathrm{spin}$ for $0.5M_\odot$, $0.7M_\odot$ and $1.0M_\odot$ stars with $P_\mathrm{spin,\,initial}/\,\mathrm{d}\in\{1,\,8\}$ and $\tau_\mathrm{dl}=3\,$Myr for different values of $f_2$. The other free parameters have been set to $f_\mathrm{shear}=0.1$ and $\delta=1$. The shaded regions of different colours denote $P_\mathrm{spin}\leq P_\mathrm{crit}\equiv \tau_\mathrm{c}/f_2$, where the colours of the shades denote the choice of $f_2$. Each $f_2$ model should only be compared to its corresponding $P_\mathrm{crit}$.  }
\label{fig:rossby_analysis}
\end{figure}

\begin{figure}
\centering
\includegraphics[width=0.4\textwidth]{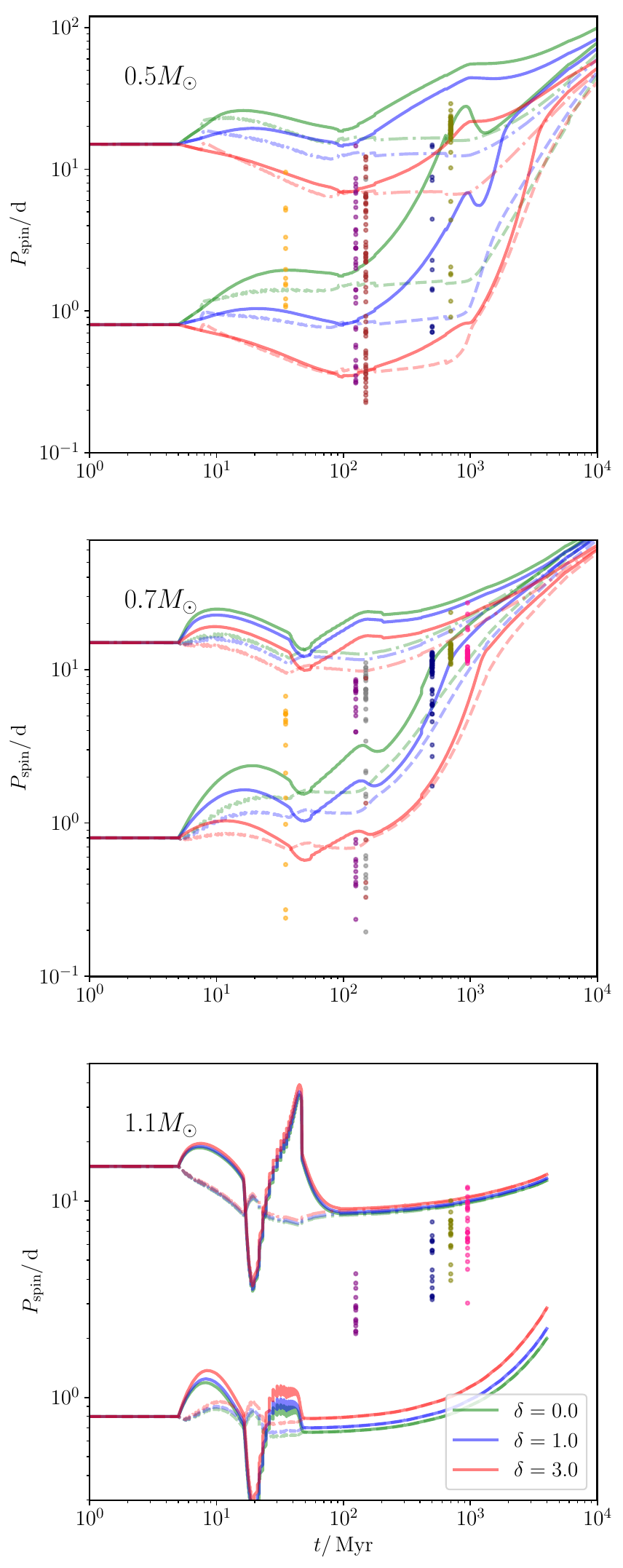}
\caption{The evolution of the core spin (dashed for the fast rotators and dash-dotted for the slow rotators) and the envelope spin (solid) $P_\mathrm{spin}$ for $0.5M_\odot$, $0.7M_\odot$ and $1.1M_\odot$ stars with $P_\mathrm{spin,\,initial}/\,\mathrm{d}\in\{1,\,8\}$ and $\tau_\mathrm{dl}=5\,$Myr for various $\delta$. The other free parameters are set to $f_\mathrm{shear}=0.1$ and $f_2=2$. The vertical arrays of dots denote observations of spins of stars in OCs of known ages by \protect\cite{2021ApJS..257...46G}. NGC2547 (35 Myr old), Pleiades (125 Myr old), M50 (150 Myr old), NGC2516 (150 Myr old), M37 (500 Myr old), Praesepe (700 Myr old) and  NGC6811 (950 Myr old) are shown as vertical arrays of dots. The observations have been allocated to a subplot if their mass $M_\ast$ lies in  $[M_\mathrm{PCD}-0.05,\,M_\mathrm{PCD}+0.05)\,M_\odot$, where $M_\mathrm{PCD}$ is the mass considered in a subplot. }
\label{fig:spot_analysis}
\end{figure}

Our MB strength depends on two uncertain free parameters, $f_2$ and $\delta$. Their effect on the MB torque can be illustrated using equation~(\ref{eq:jdot_f2_epsil2_PMS}). \textcolor{black}{Before individually analyzing these parameters, it is important to note that despite their different origin, the combined effect of $f_2$, $\delta$ and the factor $(M_\mathrm{env}/M_\ast)^2(M_\ast/M_\odot)^{-0.25}$ (equation~\ref{eq:jdot_f2_epsil2_PMS}) is to change $\dot{J}_\mathrm{w}$. As a result, there is a degeneracy in the MB evolution for certain $f_2$ and $\delta$. In the saturated regime, both $f_2$ and $\delta$ (for a given $M_\ast$) change $\dot{J}_\mathrm{w}$ by a constant factor. So the effect of reducing $f_2$ can be negated by choosing a smaller $\delta$ and vice versa. Similarly, the effect of $\delta$ can be negated by $f_2$ in the saturated regime and by using an ad hoc factor different from $(M_\ast/M_\odot)^{-0.25}$ in both regimes. It is also important to mention that it is difficult to corroborate our choices of $f_2$ and $\delta$ with observations. Our $f_2$ acts like $\chi$ in equation~(2) of \cite{Matt2015}. Their $\chi$ determines the Rossby number at which their models become unsaturated. Although there are ample observations of stellar spin, Rossby number also critically depends on the choice of $\tau_\mathrm{c}$ which is strongly model-dependent (fig.~9 of \citetalias{SYT}). The problem is even worse for $\delta$, as there is little observational data on starspot coverage in Sun-like stars, let alone that for other PCDs. This is exacerbated by the fact that stellar activity and, as a consequence, starspot coverage, is also affected by age and stellar spin (see the discussion below equation~\ref{eq:epsilon}). So, we illustrate how our tracks behave for different $f_2$ and $\delta$. Our choices of $f_2$ and $\delta$ are not observationally motivated; we vary them to assess whether we can extract valuable information from our models despite their significant uncertainty.}

We illustrate the effect of altering the MB strength on the spin trajectories by first changing $f_2$ while keeping $\delta$ fixed (Fig.~\ref{fig:rossby_analysis}). Changing $f_2$ leads to a change in $P_\mathrm{crit} = \tau_\mathrm{c}/f_2$, below which the MB torque is saturated ($\Dot{J}\propto \Omega$) and above which it is unsaturated ($\Dot{J}\propto \Omega^3$). Lower $f_2$ leads to a larger $P_\mathrm{crit}$ and also a weaker MB strength in the unsaturated regime ($\Dot{J}\propto f_2^{2}$). In Fig.~\ref{fig:rossby_analysis}, for $f_2=1$ each star shows an initial spin-up because the MB torque is too weak to counter the contraction-driven spin-up. The assumption of $f_2=2$ results in a very strong torque even in the saturated regime. This dominates over contraction until about 20~Myr for $0.5M_\odot$ and about 5~Myr for $1.0M_\odot$. For a larger $f_2$, the combined effect of a lower $P_\mathrm{crit}$ and a stronger torque in the saturated regime lead to the star attaining a larger $P_\mathrm{env,spin}$ at all times. A smaller $f_2$ leads to a shorter time to reach core-envelope corotation in all the stars. This can be understood with expression~(\ref{eq:Iineq}). The core can catch up with the envelope if it spins down more slowly. This is easier because a lower $f_2$ leads to a lower MB torque causing $-\Dot{J}_\mathrm{w}/\Dot{J}_\mathrm{shear}$ to be smaller. Changing $f_2$ affects the MB strength in the saturated regime while $\delta$ affects it at all times. Owing to our definition of equation~(\ref{eq:epsilon}), a larger $\delta$ leads to a lower MB torque for $M_\ast<1M_\odot$ and vice versa. This is illustrated in Fig.~\ref{fig:spot_analysis}, where we keep $f_2$ fixed and vary $\delta$. \textcolor{black}{Note that here we do not show a $1.0M_\odot$ star because owing to equation~(\ref{eq:epsilon}), changing $\delta$ does not affect starspot coverage for a $1.0M_\odot$ star. The model with $\delta=0$ corresponds to when the starspot coverage is independent of stellar mass. As expected, this drives the strongest spindown in stars less massive than $1M_\odot$ and the weakest spindown for more massive stars. In stars less massive than $1M_\odot$, a larger $\delta$ has an effect quite similar to a smaller $f_2$; the MB torque reduces causing a stronger initial spin-up in the PMS phase, followed by a weaker spin-down. This affects shear similarly to a lower $f_2$, because the core can catch up with the envelope if its spin-down rate is weaker. To provide a sense of how altering the MB strength matches observations, we plot} vertical arrays of dots which are observations of stars in open clusters of known ages from \cite{2021ApJS..257...46G}.


Figs~\ref{fig:rossby_analysis} and \ref{fig:spot_analysis} show that, although the envelope spins are affected by MB, the core-envelope behaviour remains qualitatively the same for the $0.5M_\odot$ and $0.7M_\odot$ stars, and the $1.0M_\odot$ and $1.1M_\odot$ stars achieve near corotation at almost the same time for all $f_2$ and $\delta$. This implies that the core-envelope behaviour is robust to changes in the efficiency of shear and magnetic braking strength. We also see that it strongly depends on the structural properties of the star and weakly depends on the initial spin. This has strong effects on the core-envelope convergence time-scale that we find in Section~\ref{subs:internal_rot}. 

\section{Results and Discussion}
\label{sec:results}
We present the results of our model addressing theoretical shortcomings and observational consequences for the evolution of low-mass solar-like stars.

\subsection{Surface rotation}
\label{subs:rotation_results}
\begin{figure*}
\centering
\includegraphics[width=1.0\textwidth]{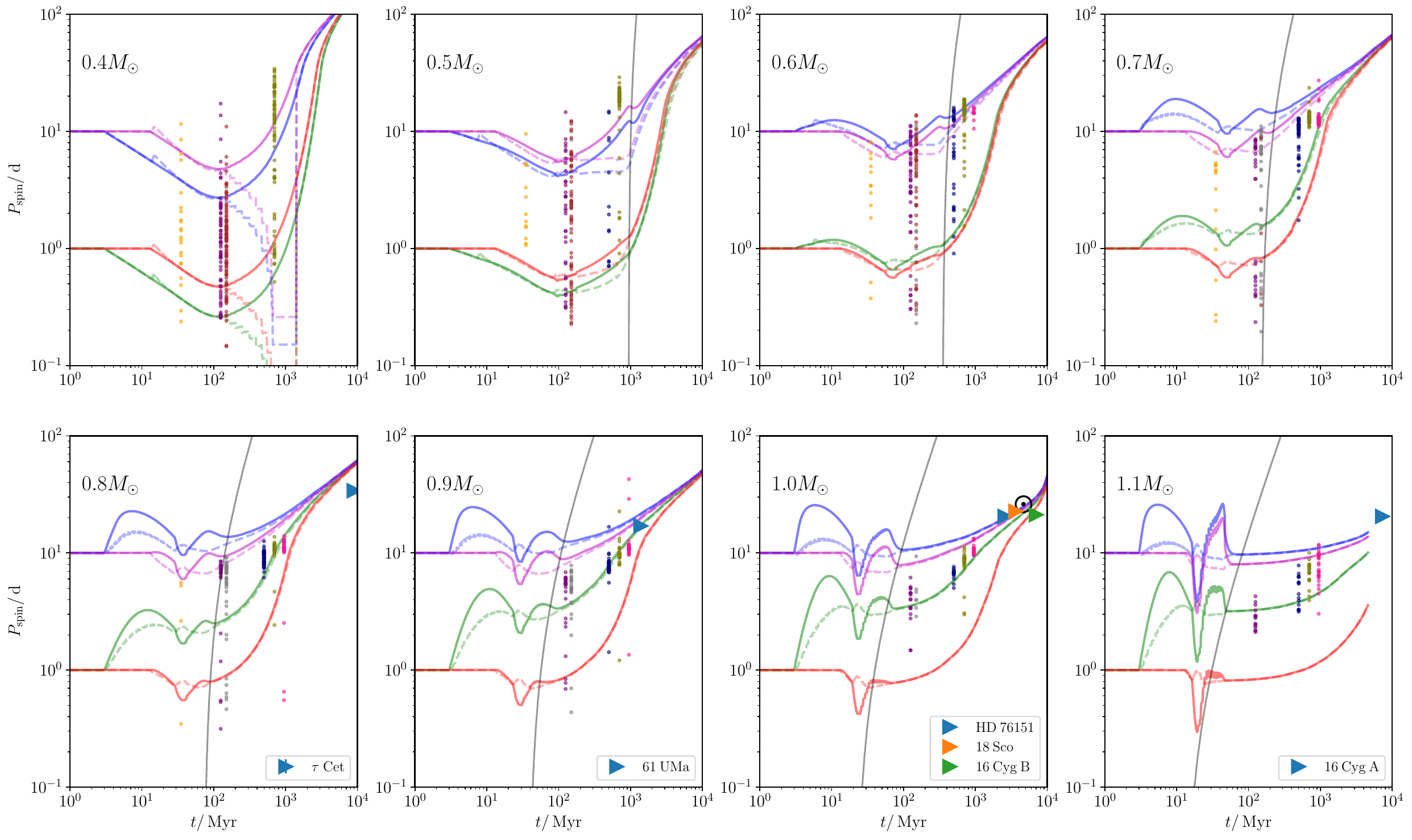}
\caption{The core (dashed) and envelope (solid) spins of PCDs of various masses with $P_\mathrm{spin,\,initial}/\,\mathrm{d}\in\{1,\,10\}$ and $\tau_\mathrm{dl}\in\{3,\,13\}\,$Myr, $f_2=2$, $f_\mathrm{shear}=0.3$ and $\delta=3$, The observations of OCs by \protect\cite{2021ApJS..257...46G} are the same as in Fig.~\protect\ref{fig:spot_analysis}. The triangles of various colours are stars reported by \protect\cite{Metcalfe2022, Metcalfe2023}. \textcolor{black}{The grey curves in the subplots from $0.5M_\odot$ to $1.1M_\odot$ are our core-envelope convergence timescale $\tau_\mathrm{converge}$ estimates from equation~(\ref{eq:tau_converge}).}  }
\label{fig:holistic_analysis}
\end{figure*}

In Fig.~\ref{fig:holistic_analysis} we plot the spin-evolution with time of the core and the envelope of PCDs with masses between $0.4M_\odot$ and $1.1M_\odot$ for different initial $P_\mathrm{spin}$ and $\tau_\mathrm{dl}$. The vertical arrays of dots are observations of OC stars spins from \cite{2021ApJS..257...46G} and triangles of various colours denote the spins and inferred ages from \cite{Metcalfe2022, Metcalfe2023}. \textcolor{black}{The free parameters $f_2=2$ and $\delta=3$ ensure a good agreement between the surface spins our models and the observations of \citet[see Fig.~\ref{fig:spot_analysis}]{2021ApJS..257...46G} for PCDs for different stellar masses, although we note that other combinations of $f_2$ and $\delta$ may lead to a similar agreement (see Section~\ref{subss:mb}). It is also important to note that the shear at the core-envelope boundary depends strongly on the chosen MB strength and weakly on $f_\mathrm{shear}$, as shown in Figs~\ref{fig:shear_analysis}, \ref{fig:rossby_analysis} and \ref{fig:spot_analysis} where for the same $f_\mathrm{shear}$ shear is minimized for a weaker MB and vice versa. To confirm this, we plot the same figure as Fig.~\ref{fig:holistic_analysis} but with $f_\mathrm{shear}=0.1$ in Fig.~\ref{fig:holistic_analysis_smallshear}. It is seen that although there is a larger difference in core-envelope spins, the qualitative behaviour of the core and envelope remains similar. More importantly, the envelope spins are virtually the same as in Fig.~\ref{fig:holistic_analysis}. In Figs~\ref{fig:holistic_analysis} and \ref{fig:holistic_analysis_smallshear} shear is automatically reduced if we ensure a good agreement between envelope spins and observations. We set $f_\mathrm{shear}=0.3$ for further analyses because it better enables us to visualise core-envelope convergence (Section~\ref{subs:internal_rot}). } 

In the mass interval $0.3M_\odot\lesssim M_\ast\lesssim0.4M_\odot$, the radiative cores of our stars vanish.  In $0.4M_\odot$ stars it disappears after 1.5~Gyr, while in $0.3M_\odot$ stars it exists for only a few Myr. At higher masses, the mass of the radiative core becomes larger and persists permanently. We do not show the spin evolution of more massive PCDs because it does not change considerably beyond their PMS contraction phase. This is due to a very weak MB torque operating in such stars (equation~\ref{eq:jdot_f2_epsil2_PMS}).

\subsubsection{M-dwarfs}
Our 
$t-P_{\rm spin}$ tracks for partially convective M-dwarfs along with those for fully convective M-dwarfs in Fig.~\ref{fig:holistic_analysis_fcmd} show that the most massive fully convective M-dwarfs ($M_\ast\approx0.35M_\odot$) are always spinning more slowly than their less massive counterparts ($M_\ast\lesssim0.3M_\odot$) as well as more massive but partly convective M-dwarfs ($0.4M_\odot\lesssim M_\ast\lesssim 0.5M_\odot$). They also spin down more within the Galactic Age than partly convective M-dwarfs \citep{2023NatAs.tmp..233L} and less massive FCMDs \citep{2023AJ....166...63J}. In our model this is due to the MB torque being a function of $M_\ast$ and $M_\mathrm{env}/\,M_\ast$ such that it attains a maximum at $0.35M_\odot$ which is the FCMD-PCD boundary. 

\subsubsection{K-dwarfs}
\label{sss:k}
K-dwarfs ($0.55M_\odot\lesssim M_\ast\lesssim 0.9M_\odot$) and the early M-dwarf of $0.5M_\odot$ show an interesting feature. It can be seen that the slow-rotating stars (initial $P_\mathrm{spin}=10\,$d) show a change in their spin-down behaviour by either temporarily halting spin-down or slightly spinning up for $P_\mathrm{spin}$ between $10$ and $20\,$d. For the $0.5M_\odot$ star, this happens between about 700~Myr to about 1.5~Gyr. This can be understood from Fig.~\ref{fig:analysis}. At around 1~Gyr, $I_\mathrm{core}$ decreases slightly, and so shear drives the spindown of the core more effectively. The strong angular momentum loss from the core feeds into the envelope, causing it to briefly spin up, thus generating a kink in $P_\mathrm{spin}$ at the surface. This effect is more prominent for larger values of $f_\mathrm{shear}$ (also see the $0.5M_\odot$ subplot in Fig.~\ref{fig:shear_analysis}). The $0.6M_\odot$ star shows the same from about 300~Myr to 500~Myr, although the effect is weaker. The extent of spin-up depends on the MB strength but its duration seems to be independent of it (see Figs~\ref{fig:rossby_analysis} and \ref{fig:spot_analysis}). The fast rotators (initial $P_\mathrm{spin}=1\,$d) also show a change in their spin-down rates at the same time for a given PCD. The stalling effect is not visible in stars more massive than about $0.8M_\odot$ because the core and the envelope converge in their spins quite early on. This fits with observations that show that K-dwarfs experience a mass-dependent epoch of stalled spin-down and that the duration of stalling is longer for less-massive stars \citep{2019ApJ...879...49C}. \cite{Curtis2020} and \cite{Spada2020} stated that an internal redistribution of angular momentum, owing to core-envelope coupling, can explain this behaviour. However, our models show that simply altering the internal angular momentum transfer efficiency has a very weak effect on the spin tracks of PCDs (Fig.~\ref{fig:shear_analysis}). On the other hand, even decoupled tracks in Figs~\ref{fig:rossby_analysis} and \ref{fig:spot_analysis} show this stalling at the same age for a given star, albeit the $P_\mathrm{spin}$ at which it occurs is MB-dependent. This suggests that the stalling of $P_\mathrm{spin}$ in slow rotators is likely a MB-dependent and definitely a stellar-structure-dependent phenomenon (see the $P_\mathrm{spin}$ and $I$ behaviour in Fig.~\ref{fig:analysis}).

\subsubsection{G-dwarfs}

It is seen that there is a weakening in the spin-down of both the slow- and the fast-rotators of the $1.0M_\odot$ and the $1.1M_\odot$ stars owing to a very weak MB torque acting on the envelope. This fits with the predictions of \cite{vanSaders2019} who stated that a weak MB operates in slowly rotating $M_\ast\gtrsim1M_\odot$ stars. This also explains the long-period pileup of such stars beyond 1~Gyr for several Gyr reported by \cite{David2022}. For the fast-rotating $1M_\odot$ star in Fig.~\ref{fig:holistic_analysis} (green track), we see a rapid spin-down between about 100~Myr to 800~Myr when $P_\mathrm{spin}$ goes from a few to about 10~d while $P_\mathrm{spin}$ barely changes for the slow rotators. This transition region can explain the pileup at the short-period edge in the results of \cite{David2022} and is likely a dead-zone-dependent effect (fig.~4 of \citetalias{SYT}). However, this transition is not likely related to the core-envelope coupling mechanism in such stars because this transition is independent of the efficiency of shear (Fig.~\ref{fig:shear_analysis}). Our tracks also show that such stars have corotating cores and envelopes at about 50~Myr. This is earlier than the observed pileups at a few Gyr. The strength of this rapid spin-down in fast rotators depends on MB (Figs~\ref{fig:rossby_analysis} and \ref{fig:spot_analysis}). The age at which this transition takes place is the same for a given star. This suggests that it is also a stellar-structure-dependent phenomenon.

\subsection{Internal rotation}
\label{subs:internal_rot}
The tracks in Fig.~\ref{fig:holistic_analysis} (and Fig.~\ref{fig:holistic_analysis_smallshear}) show that the cores and envelopes achieve near corotation at a time that depends on their mass and spins. More massive stars achieve corotation earlier and rapidly spinning stars achieve corotation earlier than their slowly rotating counterparts of the same mass. The core-envelope coupling timescale $\tau_\mathrm{couple}$ determines when the core and the envelope achieve corotation. It has been an important variable affecting the evolution of PCDs \citep{Denissenkov2010, Eggenberger2019, Spada2020, Btrisey2023}. However, Figs~\ref{fig:shear_analysis}, \ref{fig:rossby_analysis} and \ref{fig:spot_analysis} show that $\tau_\mathrm{couple}$  depends on the efficiency of shear and the behaviour of the MB torque. So we define $\tau_\mathrm{converge}$ as the age at which the core and the envelope achieve near corotation for fast rotators and tend to achieve corotation for slow rotators. This time weakly depends on uncertainties in MB and internal angular momentum redistribution efficiency\footnote{The slowly rotating $0.5M_\odot$ and $0.7M_\odot$ stars show a modest increase in their differential rotation as their core and envelope spins converge for certain choices of $f_\mathrm{shear}$, $f_2$ and $\delta$. This is alleviated if we choose a larger $f_\mathrm{shear}$ for a given MB strength or a lower MB strength for a given $f_\mathrm{shear}$. }. \textcolor{black}{This is illustrated in Figs~\ref{fig:shear_analysis}, \ref{fig:rossby_analysis} and \ref{fig:spot_analysis}, where the core and the envelope commence their convergence at about the same time. For weaker MB or stronger shear efficiency, the convergence is stronger and vice versa.} This time can be evaluated from the abrupt drop in $\Delta\Omega_\mathrm{shear}/\,\Omega_\mathrm{env}$ after a local maximum in $\Delta\Omega_\mathrm{shear}/\Omega_\mathrm{env}$ (Fig.~\ref{fig:holistic_diffrot_analysis}). It can also be seen in Fig.~\ref{fig:holistic_analysis} where, for the $0.5M_\odot$ star, the core and the envelope spins rapidly converge at about 1~Gyr. For the $1M_\odot$ star this happens at about 50~Myr. \textcolor{black}{Because the convergence is seen better in Fig.~\ref{fig:holistic_analysis} compared to that in Fig.~\ref{fig:holistic_analysis_smallshear},} we use the tracks from this figure to find an empirical expression for $\tau_\mathrm{converge}$ that depends on both $M_\ast$ and $P_\mathrm{spin}$. It is given by
\begin{center}
\begin{equation}
\label{eq:tau_converge}
\displaystyle{\frac{\tau_\mathrm{converge}}{\mathrm{Myr}} \approx 22\left(\frac{M_\ast}{M_\odot}\right)^{-5.42}+ 17.82\left(\frac{P_\mathrm{spin}}{\mathrm{d}}\right)^{0.59}}.
 \end{equation}
\end{center}
 This is valid for $0.5\lesssim M_\ast/M_\odot\lesssim1.1$ and at all periods above breakup. \textcolor{black}{We plot $\tau_\mathrm{converge}$ as grey curves in Figs~\ref{fig:holistic_analysis} and \ref{fig:holistic_analysis_smallshear}. Note that, although equation~(\ref{eq:tau_converge}) is for the set of free parameters used for Fig.~\ref{fig:holistic_analysis}, it provides $\tau_\mathrm{converge}$ still in general agreement with the core-envelope convergence time of the models shown in Fig.~\ref{fig:holistic_analysis_smallshear} for which the shear efficiency is weaker.} Fig.~\ref{fig:holistic_analysis} shows that for fast rotators $\tau_\mathrm{converge}\approx\tau_\mathrm{couple}$ while slow rotators have an ever-decreasing differential rotation beyond $\tau_\mathrm{converge}$. Interestingly, equation~(\ref{eq:tau_converge}) is similar to the expression for $\tau_\mathrm{couple}$ obtained by \cite{Spada2020} using the semi-empirical fitting of \cite{2015A&A...584A..30L}, given by
 \begin{center}
\begin{equation}
\label{eq:tau_coup_spada}
\displaystyle{\frac{\tau_\mathrm{couple}}{\mathrm{Myr}} = 22\left(\frac{M_\ast}{M_\odot}\right)^{-5.6}}.
 \end{equation}
\end{center}
It is seen that equation~(\ref{eq:tau_converge}) tends to equation~(\ref{eq:tau_coup_spada}) as $P_\mathrm{spin}\xrightarrow[]{}0$. That is, our expression reproduces the $\tau_\mathrm{couple}$ derived by \cite{2015A&A...584A..30L} for fast rotators. 



\subsection{Other observables}
\label{subs:other_obs}

\begin{figure*}
\centering
\includegraphics[width=1.0\textwidth]{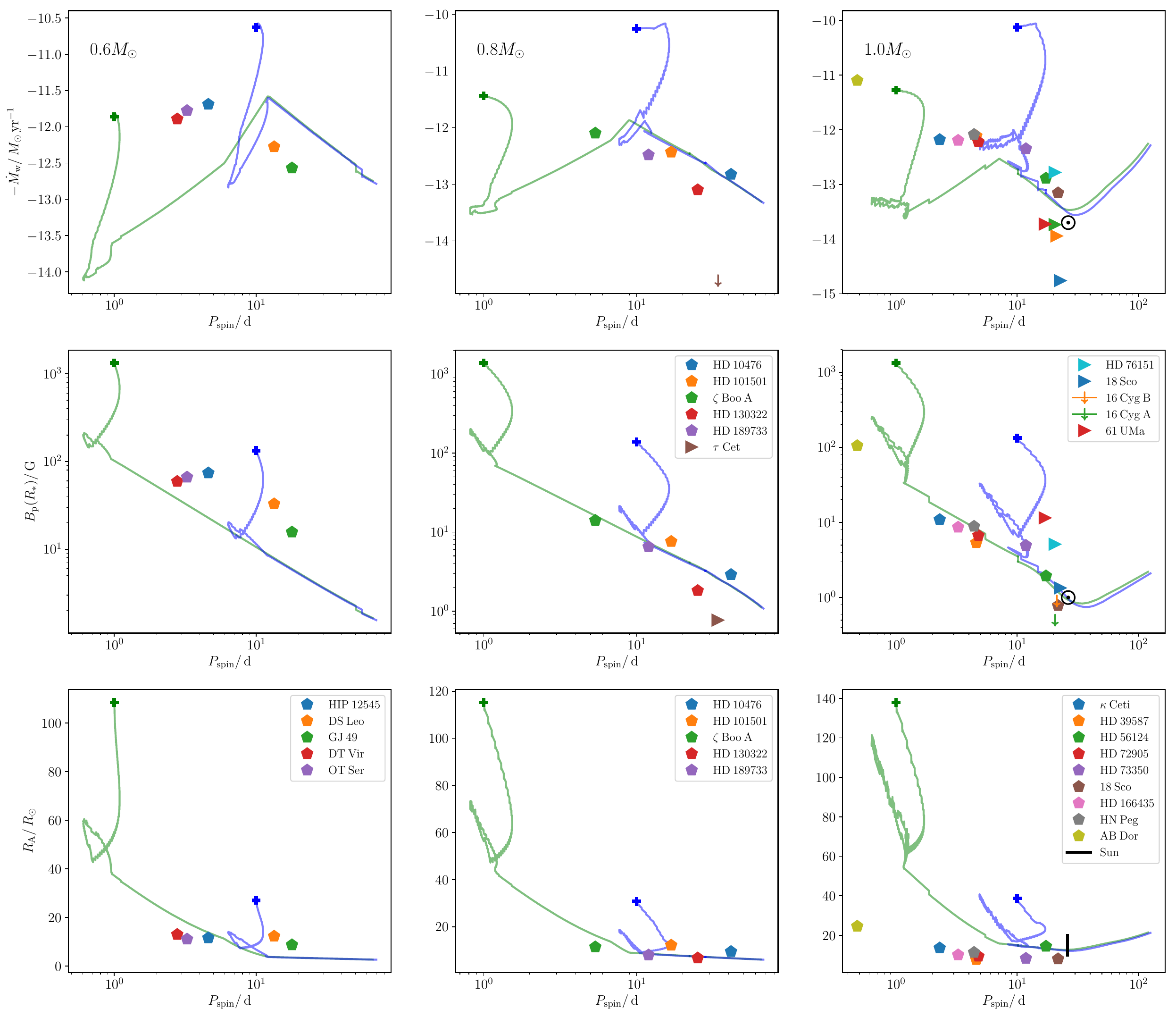}
\caption{$P_\mathrm{spin}$-dependent evolution of wind mass loss $\Dot{M}_\mathrm{w}$, dipole component of the poliodal magnetic field at the surface $B_\mathrm{p}(R_\ast)$ and the Alfvén radius $R_\mathrm{A}$ for trajectories with initial $P_\mathrm{spin}\in\{1,\,10\}$d, $\tau_\mathrm{dl}=5\,$Myr and free parameters set to $f_2=2$, $f_\mathrm{shear}=0.3$, $\delta=3$. The wiggles of the tracks at early times are because $\Delta\Omega_\mathrm{shear}/\Omega_\mathrm{env}$ is stiff for PCDs (also see Fig.~\ref{fig:holistic_diffrot_analysis}). 
The pluses of the same colour indicate the starting points of the trajectories. The observations are from \protect\citet[pentagons]{2018MNRAS.475L..25V} and \protect\citet[triangles]{Metcalfe2022, Metcalfe2023}. The minima in the $\Dot{M}_\mathrm{w}$-, $B_\mathrm{p}(R_\ast)$- and $R_\mathrm{A}$-$P_\mathrm{spin}$ planes at about 30~d in the $1M_\odot$ star are due to its expansion after terminal-age main sequence at about 8~Gyr. }
\label{fig:holistic_otherobs}
\end{figure*}

As illustrated by \citetalias{SYT}, a unique advantage of our model is that we obtain analytical estimates of the winds, the dipole component of the magnetic field (hereinafter just magnetic field) and the Alfvén radius at all times for a given star (equations~\ref{eq:mlconv}, \ref{eq:bpconv} and \ref{eq:ra}). These are plotted for several PCDs in Fig.~\ref{fig:holistic_otherobs}. The observationally inferred parameters are from \citet[pentagons]{2018MNRAS.475L..25V} and \citet[triangles]{Metcalfe2022, Metcalfe2023}. It is seen that, in general, our estimates are the best for $1M_\odot$ stars. This is possibly because we used solar data to calibrate our magnetic field (by setting $f_1=0.115$) and MB torque (equations~\ref{eq:mlconv0} and \ref{eq:epsilon}). The estimates are not bad for $0.8M_\odot$ stars either. \textcolor{black}{Interestingly, in the $\dot{M}_\mathrm{w}-P_\mathrm{spin}$ plane $\Dot{M}_\mathrm{w}$ drops to $\approx \Dot{M}_\odot$ when $P_\mathrm{spin}\approx 1$~d early on for the $1M_\odot$ star evolved with initial $P_\mathrm{spin}=1\,$d (the green track).} This illustrates that winds as strong as that of the present-day Sun are possible in young solar-like stars. This can explain the observation reported by \cite{2014ApJ...781L..33W} that found $\Dot{M}_\mathrm{w}=0.5\Dot{M}_\odot$ in a young, rapidly rotating solar analog star $\pi^1$~UMa. 

It is seen that our estimates deviate for rapidly spinning configurations for $R_\mathrm{A}$ (a factor of a few) and $\Dot{M}_\mathrm{w}$ (about an order of magnitude). This suggests that $f_\mathrm{corot}$ in equation~(\ref{eq:fcorot}) may be poorly defined. Overall, the surface magnetic field $B_\mathrm{p}(R_\ast)$ estimates seem to be most robust (within a factor of a few) once the star begins to spin down smoothly. So we derive an empirical expression for this. Our expression for $B_\mathrm{p}(R_\ast)$ (equation~\ref{eq:bpconv}) is a simple function of stellar parameters $M_\ast$, $L_\ast$ and $R_\ast$ provided we know the spin and $\rho_\mathrm{env}$. So, using equations~(\ref{eq:vc}), (\ref{eq:gamma}) and (\ref{eq:bpconv}), we can write the magnetic field for any star $B_\mathrm{p\ast}$ as

\begin{center}
\begin{equation}
\label{eq:bp_fit_0}
\displaystyle{\frac{B_\mathrm{p}}{B_\mathrm{p\odot}} = \left(\frac{L_\ast}{L_\odot}\right)^{1/3} \left(\frac{M_\ast}{M_\odot}\right)^{5/6} \left(\frac{R_\ast}{R_\odot}\right)^{7/6} \left(\frac{\Omega_\ast}{\Omega_\odot}\right) \left(\frac{\rho_\mathrm{env}}{\rho_\mathrm{env\odot}}\right)^{1/2}  },
 \end{equation}
\end{center}
where we obtain $\rho_\mathrm{env\odot}= 0.069\, \mathrm{g\,cm^{-3}}$ from our STARS models. We fit $(\rho_\mathrm{env}/\rho_\mathrm{env\odot})^{0.5}$ as a function of age $t$ and $M_\ast$ for $0.8M_\odot$, $0.9M_\odot$, $1.0M_\odot$ and $1.1M_\odot$ stars, obtaining 

\begin{center}
\begin{equation}
\label{eq:rhoenv_fit}
\displaystyle{\left(\frac{\rho_\mathrm{env}}{\rho_\mathrm{env\odot}}\right)^{1/2} \approx -0.12\left(\frac{t}{\mathrm{Gyr}}\right) -9.6 \left(\frac{M_\ast}{M_\odot}\right) + 11.6  }.
 \end{equation}
\end{center}
Using $L_\ast \propto M_\ast^{3.9}$ and $R_\ast \propto M_\ast^{0.8}$ for main-sequence stars, we obtain
\begin{center}
\begin{equation}
\label{eq:bp_fit}
\displaystyle{\frac{B_\mathrm{p}}{B_\mathrm{p\odot}} \approx \left(\frac{M_\ast}{M_\odot}\right)^{3.07} \left(\frac{\Omega_\ast}{\Omega_\odot}\right) \left(-0.12\left(\frac{t}{\mathrm{Gyr}}\right) -9.6\left(\frac{M_\ast}{M_\odot}\right) + 11.6\right)}.
 \end{equation}
\end{center}
This expression gives a simple estimate of the surface magnetic field within a factor of a few for solar-like main-sequence stars.

\subsubsection{Is shear causing Praesepe K-dwarfs to be overactive?}
It was recently reported by \cite{Cao2023} that slowly rotating ($P_\mathrm{spin} \approx 10\,$d) K-stars in the approximately 700~Myr-old Praesepe OC are unusually overactive. The observation of enhanced activity in such stars, together with an apparent stalling of spin-down between the ages of Praesepe and the 950~Myr-old NGC~6811 \citep{2019ApJ...879...49C} is explained with core-envelope decoupling leading to shear-enhanced activity in the convective envelope as well as a brief stalling of spin-down. 

In Fig.~\ref{fig:holistic_analysis}, it is seen that all $0.5M_\odot$ stars achieve near corotation between the ages of Praesepe and NGC~6811 while the $0.7M_\odot$ stars do so much earlier at about 200~Myr. Beyond this age, the spinning-down envelope drags the core with increasing efficiency, thereby reducing differential rotation. The amount of differential rotation predicted by our model is negligible compared to that by \cite{Cao2023} \textcolor{black}{even if we consider the models from Fig.~\ref{fig:holistic_analysis_smallshear} with a larger overall difference in the core-envelope spins (owing to a smaller $f_\mathrm{shear}$). At 700~Myr our slow-rotating tracks for the $0.7M_\odot$ and $0.8M_\odot$ stars have $P_\mathrm{spin,\,env}=xP_\mathrm{spin,core}$, where $x<2$, while $P_\mathrm{spin,\,env}\approx 5P_\mathrm{spin,core}$ for a $0.75M_\odot$ star in fig.~2 of \cite{Cao2023}.} We have discussed in Section~\ref{sss:k} that the stalling of spin-down is a structure- and MB-dependent phenomenon and its duration and occurrence are sensitive to MB uncertainties. In addition, Fig.~\ref{fig:holistic_otherobs} shows that the $0.6M_\odot$ stars experience a sharp increase in their wind mass-loss rate $\Dot{M}_\mathrm{w}$ at $P_\mathrm{spin}$ of about 10~d and that more massive stars do this at slower spins with the effect being less pronounced. This is due to the transition of the star from the saturated regime to the unsaturated regime (equation~\ref{eq:mlconv}). Using $\Dot{M}_\mathrm{w}$ as a proxy for stellar activity \citep{2016MNRAS.458.1548B}, we argue that this abrupt spike in wind mass loss in an otherwise monotonically decreasing magnetic field \textcolor{black}{(albeit the initial evolution owing to structural changes and small wiggles)} and Alfvén radius profile may explain the unusual activity of such stars and that it is also a structure- and MB-dependent effect. \textcolor{black}{However, the envelope $P_\mathrm{spin}$ at which this peak appears depends strongly on the choice of $f_2$, which is a free parameter. We discuss in Section~\ref{subs:rotation_results} that our choice of $f_2$ and $\delta$ is motivated by the fact that our $P_\mathrm{spin}$ tracks corroborate with observations, but other choices that lead to a similar agreement may change the strength and position of this peak in our tracks.}

\textcolor{black}{Finally, we note that Praesepe has non-solar metallicity, because of which its constituent K-stars may behave differently to our solar metallicity stars. Although metallicity was shown to have a negligible effect on the evolution of M-dwarfs (fig.~A.1 of \citetalias{SYT}), we have not undertaken such a study for more massive stars. However, \cite{1998A&A...337..403B} show negligible differences between the evolution of solar-like stars of different metallicities. Such a study will be considered in the future.}

\subsection{The Faint Young Sun paradox}
\label{subs:youngsun}

\begin{figure}
\includegraphics[width=0.45\textwidth]{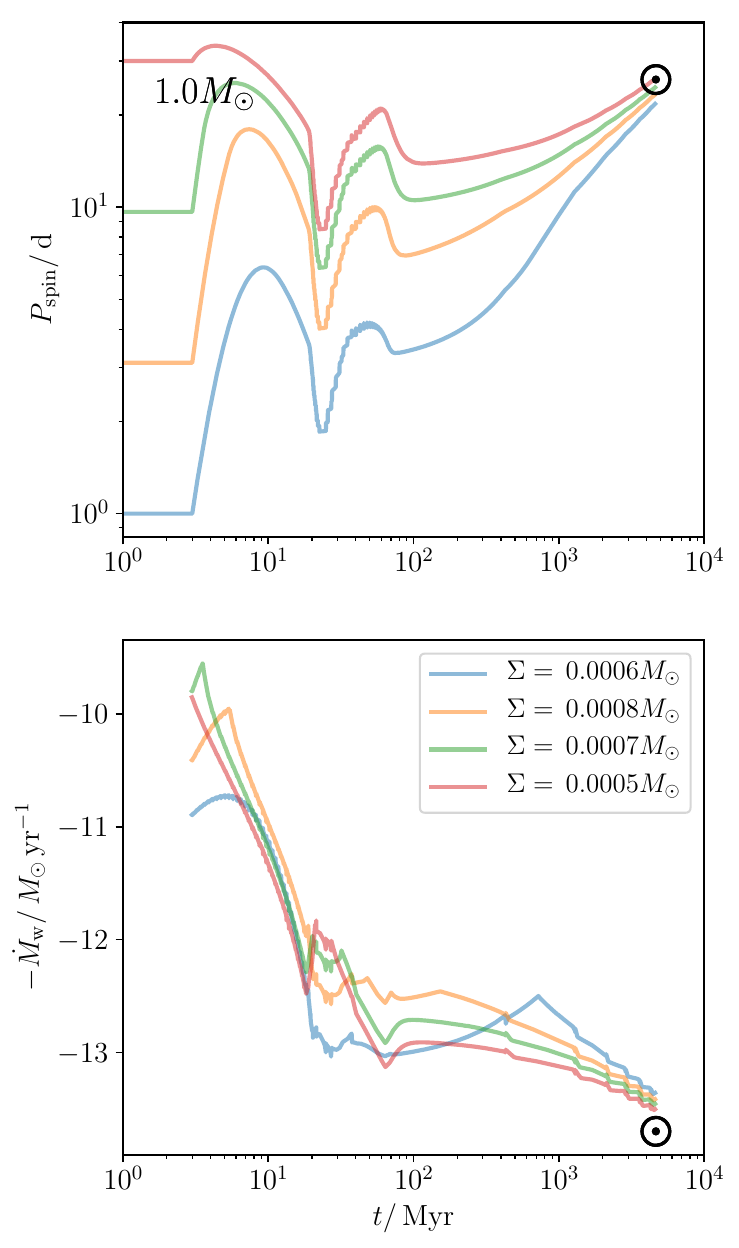}
\caption{The evolution of the surface spin $P_\mathrm{spin}$ and mass-loss rate $\Dot{M}_\mathrm{w}$ with time for a $1M_\odot$ star with $\tau_\mathrm{dl}=3\,$Myr, initial $P_\mathrm{spin}\in [1,\,30]$d equally spaced logarithmically. The free parameters are the same as in Fig.~\ref{fig:holistic_otherobs} and $\Sigma$ denotes the total mass lost between $\tau_\mathrm{dl}$ and $t_\odot$.}
\label{fig:youngsun}
\end{figure}

When the Sun was formed about 4.6 Gyr ago as a zero-age main-sequence star, it had a luminosity of about $0.7L_\odot$. Based on this, calculations of the Earth's ancient climate, under certain assumptions of its albedo and atmospheric composition, yield an average terrestrial temperature that is below water's freezing point. However, there is ample geological evidence to suggest that water on the early Earth's surface was liquid. This apparent contradiction is known as the Faint Young Sun paradox \citep{Feulner2012}. Explanations for this problem include greenhouse gas solutions \citep{Pavlov2000}, cosmic rays \citep{2003JGRA..108.1437S}, albedo effects \citep{2010Natur.464..744R} etc. 

One of the solutions to this problem is the massive young Sun hypothesis \citep{Minton2007}. This theory suggests that when the Sun was born it was more massive than $1M_\odot$. As a result its luminosity was modestly higher and the Earth was closer to it. This kept Earth warm enough for liquid water. Over time, strong solar winds reduced the Sun's mass to what we observe today while its luminosity increased. The reduction in stellar mass balanced the luminosity increase so as to have consistently warm enough temperatures on Earth to keep water liquid.  \cite{Minton2007} calculated the minimum mass loss that our Sun would have had to experience to resolve this paradox and found it to be about $0.026M_\odot$ and that the Sun would need to have remained this for about 2~Gyr after its formation. With our detailed mass loss model we can test this hypothesis. We calculate the mass-loss and spin evolution for a $1.0M_\odot$ star. Because our model does not alter the stellar properties due to mass loss, we cannot track the evolution of the mass loss in response to stellar properties. However, equation~(\ref{eq:mlconv}) shows that for our partly convective stars $\Dot{M}_\mathrm{w}\propto (M_\mathrm{env}/\,M_\ast)^2 M_\ast^{-1.25}$, so the larger the stellar mass, the weaker the winds that carry angular momentum. Thus, the mass-loss profile of our $1.0M_\odot$ star represents the upper limit to the overall mass loss possible. We define $\Sigma$ as the total mass lost before $t_\odot$ as

\begin{center}
\begin{equation}
\label{eq:Sigma}
\Sigma = \int_{\tau_\mathrm{dl}}^{t_\odot}\Dot{M}_\mathrm{w}\,\mathrm{d}t.
 \end{equation}
\end{center}
In order to maximize $\Sigma$, we set $\tau_\mathrm{dl}=3\,$Myr and investigate initial spins equally spaced logarithmically between 1 and 30~d (Fig.~\ref{fig:youngsun}). It is seen that $\Dot{M}_\mathrm{w}$ drops to about an order of magnitude the current solar wind mass-loss rate between 20~Myr and 80~Myr and then converges to $\Dot{M}_\odot$. However, $\Sigma$ is always smaller than 
 the $0.026M_\odot$ required to solve the faint young Sun problem. Even the most optimistic $\Sigma\approx0.001M_\odot \ll 0.026M_\odot$ by about an order of magnitude. So, if our model is correct, we conclude that the massive young Sun theory is unlikely to be a viable solution to the faint young Sun problem as suggested by \cite{Minton2007}.


\section{Conclusions}
\label{sec:conclusions}

We have built a model to evolve the spins of the cores and the envelopes of partly convective dwarf stars (PCDs, $0.35\lesssim M_\ast/M_\odot\lesssim1.3$). This work extends that of \cite{SYT} when we modelled the spin evolution of fully convective M-dwarf stars (FCMDs, $M_\ast\lesssim0.35 M_\odot$). We list the important components of our model.
\begin{enumerate}
    \item The convective envelopes and the radiative cores of our PCDs rotate as solid bodies. The spin of the core is influenced by shear, which acts to transport angular momentum between the core and the envelope. The spin of the envelope is influenced by angular momentum loss by magnetic braking (MB) and shear.
    \item Our MB formalism is the same as that of \cite{SYT} with changes that incorporate the fact that the star is partly convective (Fig.~\ref{fig:cartoon}). Our expression for the shear is from \cite{Zangrilli1997}. We have also made additional modifications to the expressions for wind mass loss and Alfvén radius which influence MB (Table~\ref{tab:table_change}). 
    \item We work with three free parameters: our $f_\mathrm{shear}$ governs the efficiency of shear; $f_2$ marks the transition of the MB torque from the saturated $(\Dot{J}\propto\Omega)$ to the unsaturated $(\Dot{J}\propto\Omega^3)$ regime; and $\delta$ governs the behaviour of the wind-emitting surface of the star. Our $f_\mathrm{shear}$ governs the strength of shear while $f_2$ and $\delta$ govern the MB strength. \textcolor{black}{These free parameters are difficult to obtain accurately, so we study the behaviour of our model to changes in them.}
    \item We use two coupled ordinary differential equations to solve for the spins of the cores and the envelopes of a PCD from the time of its disc-dispersal to the end of the main sequence or the Galactic Age, whichever is earlier.
    
\end{enumerate}
We arrive at several important conclusions. 

\begin{enumerate}
    \item Our spin trajectories depend weakly on the strength of shear, with the dependence becoming weaker as the star attains a stable configuration after the pre-main sequence contraction phase. On the other hand, our spin trajectories strongly depend on the behaviour of the MB torque. We illustrate this by changing $f_2$ and $\delta$.

    \item The most massive FCMDs ($M_\ast\approx0.35M_\odot$) experience stronger MB torques than less massive FCMDs ($M_\ast\lesssim0.35M_\odot$) and more massive PCDs (Figs~\ref{fig:holistic_analysis} and \ref{fig:holistic_analysis_fcmd}). These stars experience the strongest MB torque throughout and spin down the most within the Galactic Age.

    \item Slow-rotating stars with masses between about $0.5M_\odot$ and $0.8M_\odot$ show a mass-dependent duration of stalled spin-down or minor spin-up (Figs~\ref{fig:holistic_analysis} and \ref{fig:holistic_analysis_smallshear}) such that less massive PCDs show a longer duration of stalling and at later times. Although the spin period at which this happens depends on the MB strength (Figs~\ref{fig:rossby_analysis} and \ref{fig:spot_analysis}), its duration only weakly depends on MB and shear and is strongly stellar-structure-dependent. Our models show that a temporary epoch of stalled spin-down in K-dwarfs \citep{2019ApJ...879...49C,Curtis2020} is likely a MB- and stellar-structure dependent effect. Such a MB-dependent effect simultaneously explains why these K-dwarfs are observed to be unusually active \citep{Cao2023}. This is because in K-stars there is a maximum in the wind mass-loss rate at $P_\mathrm{spin}$ between 15 and $20\,$d which is where the MB transitions from the saturated to the unsaturated regime (Fig.~\ref{fig:holistic_otherobs} and equation~\ref{eq:mlconv}). Using wind mass loss as a proxy for stellar activity, we argue that this maximum in an otherwise smooth magnetic field and Alfvén radius profile can \textcolor{black}{possibly} explain the unusual activity of such stars. 

    \item The pileup in the spin--mass plane observed by \cite{David2022} for massive PCDs ($M_\ast\gtrsim1.0 M_\odot$) is also reflected in our trajectories of $1.0M_\odot$ and $1.1M_\odot$ stars, wherein slow rotators experience weakened MB torques after $\simeq$ Gyr time-scales. This is in accordance with the weakened magnetic braking paradigm proposed by \cite{vanSaders2019}. However, our fast rotators also transit from spin periods of a few to about 10~d. This is a dead-zone-dependent effect (fig.~4 of \citealt{SYT}). 

    \item The stalling of spin-down and unusual activity of K-dwarfs, and the pileups of G-dwarfs are unlikely to be the result of core-envelope coupling effects. For K-dwarfs, this is because the spin-stalling in our modelled stars only weakly depends on the efficiency of shear and the differential rotation required to explain the activity of K-dwarfs by shear-enhanced activity disagrees with our differential rotation estimates. The pileups of G-dwarfs cannot be explained by core-envelope coupling effects because such stars have corotating cores and envelopes at about 50~Myr, which is very early compared to the observed pileups at a few Gyr.

    \item It is difficult to define a core-envelope corotation or coupling time-scale $\tau_\mathrm{couple}$ when the core and the envelope achieve near corotation. This is because the time at which differential rotation becomes negligible in a trajectory is not only mass- and spin-dependent, but also depends strongly on MB uncertainties (Fig~\ref{fig:rossby_analysis} and \ref{fig:spot_analysis}). However, the time at which the core and the envelope commence converging to a common spin weakly depends on MB-uncertainties and shear and is a stellar-structure-dependent phenomenon. We find that the core-envelope convergence time-scale $\tau_\mathrm{converge}$ can be empirically estimated by equation~(\ref{eq:tau_converge}). This expression is valid for all PCDs and spin periods above breakup. We also find that for fast rotators $\tau_\mathrm{couple}\approx\tau_\mathrm{converge}$.

    \item We track the evolution of other observable parameters such as wind mass-loss rate, dipole component of the surface poloidal magnetic field $B_\mathrm{p}$ and the Alfvén radius and compare with observations. We find that our estimates agree better for around $1M_\odot$ stars and at longer periods. We also show that our model predicts weak, solar-like winds from young, rapidly rotating solar analogs which can explain the weak winds observed in a young, rapidly rotating solar analog star $\pi^1$ UMa. It is seen that our $B_\mathrm{p}$ estimates are the most robust. We find that these can be encapsulated within a factor of a few by an empirical expression given by equation~(\ref{eq:bp_fit}).
    
    \item We assess the validity of the massive young Sun hypothesis \citep{Minton2007} as a solution to the faint young Sun problem. This theory states that the Sun was born about 3 percent more massive than it is now, and that wind mass loss in the first few Gyr combined with luminosity increase as the Sun crosses the main sequence kept the early Earth's temperature warm enough for water to exist as liquid. We test this hypothesis with our wind mass-loss model and find that the maximum cumulative mass loss our Sun could have experienced is about $0.001M_\odot$, which is an order of magnitude lower than the minimum required to explain the problem with this theory. 
\end{enumerate}

\section*{Acknowledgements}
\textcolor{black}{We thank the anonymous referee for a detailed review, which greatly improved this work.} AS thanks the Gates Cambridge Trust for his scholarship. AS thanks Jim Fuller for suggesting to undertake this work, as well as discussions on the spin evolution of low-mass stars. AS thanks Eric Blackman for insightful discussions on possible extensions of this work. CAT thanks Churchill College for his fellowship. AS and CAT thank James Pringle for drawing our attention to the faint young Sun problem.

\section*{Data availability}
No new data were generated in support of this research. Any numerical codes and related data generated during the work will be made available whenever requested by readers.



\bibliographystyle{mnras}
\bibliography{mnras_template} 



\appendix

\section{The effect of changing the efficiency of core-envelope shear}
\label{app:weak_shear}
\begin{figure*}
\centering
\includegraphics[width=0.95\textwidth]{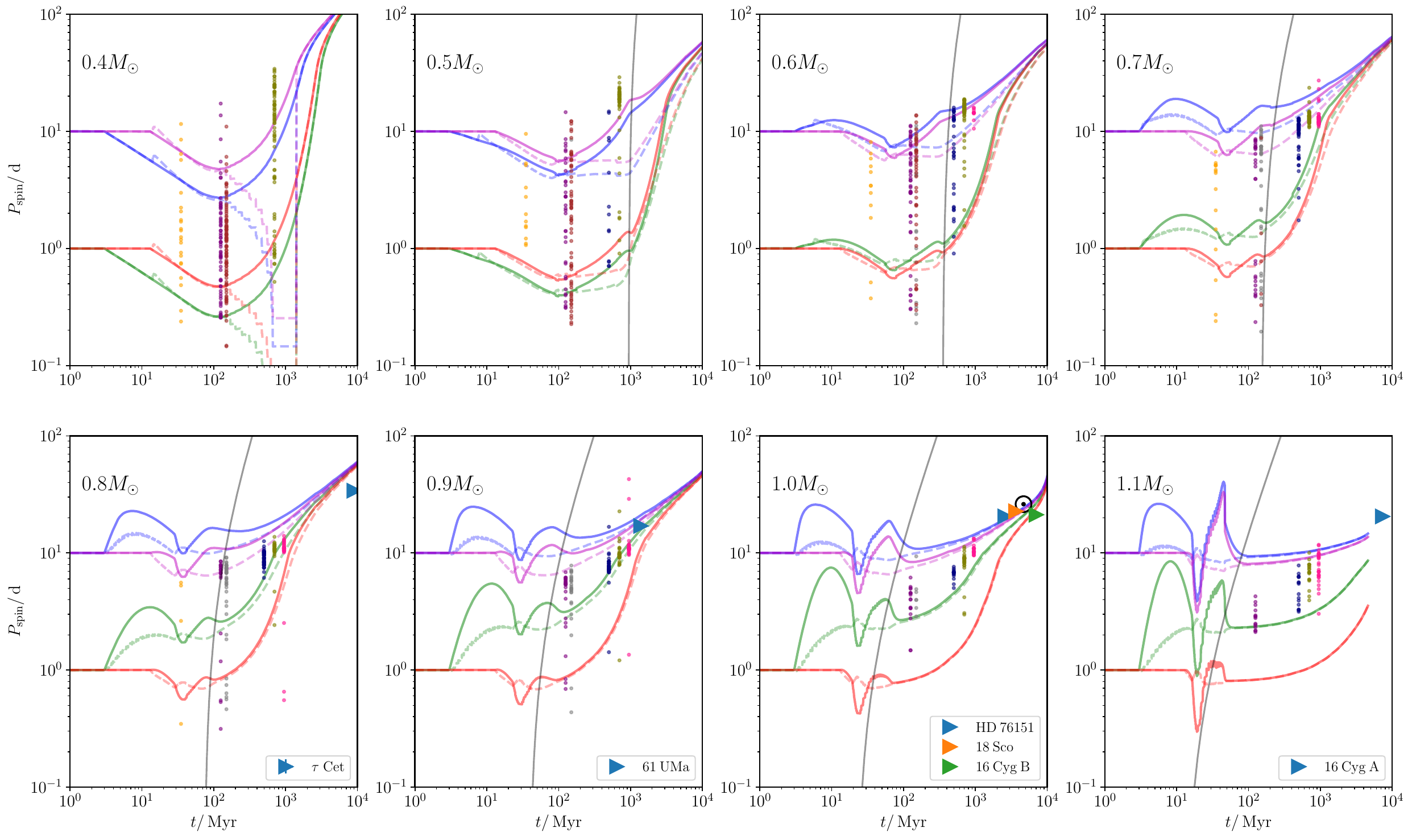}
\caption{As Fig.~\ref{fig:holistic_analysis} but with $f_\mathrm{shear}=0.1$.  }
\label{fig:holistic_analysis_smallshear}
\end{figure*}

To illustrate the weak effect of the efficiency of shear $f_\mathrm{shear}$ on our evolution tracks, we plot the same results as in Fig.~\ref{fig:holistic_analysis} but for a weaker $f_\mathrm{shear}$ in Fig.~\ref{fig:holistic_analysis_smallshear}.
\section{Evolution of fully convective M-dwarfs}
\label{app:FCMD}

\begin{figure*}
\centering
\includegraphics[width=0.95\textwidth]{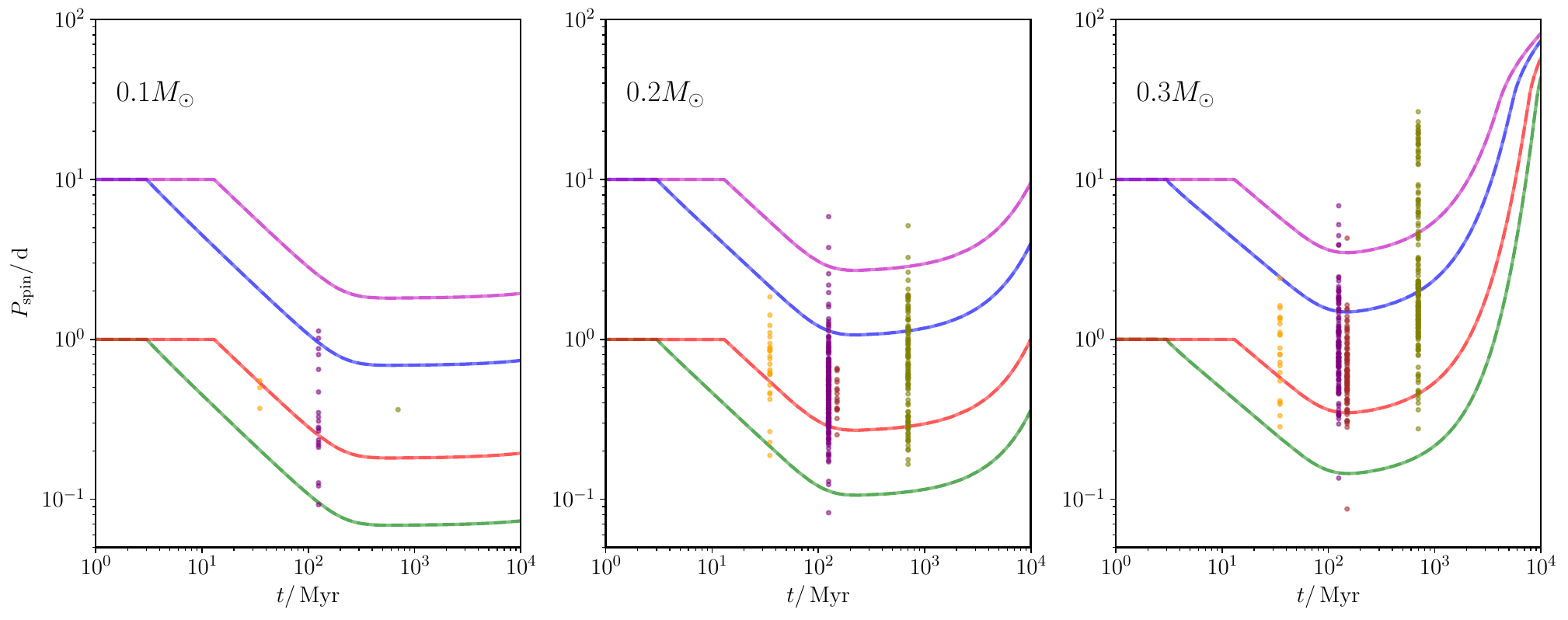}
\caption{The evolution of $P_\mathrm{spin}$ with time for FCMDs with $\tau_\mathrm{dl}\in \{3,\,13\}$~Myr, $f_2=2$, $f_\mathrm{shear}=0.3$ and $\delta=3$. The dots represent the same observations as in Fig.~\ref{fig:spot_analysis}.  }
\label{fig:holistic_analysis_fcmd}
\end{figure*}

For the sake of completeness, we plot the spin-evolution of FCMDs with our MB model in Fig.~\ref{fig:holistic_analysis_fcmd}. These stars have no radiative cores so our plots show the cores and the envelopes corotating at all times. The trajectories deviate from those of \citetalias{SYT} for low-mass FCMDs because of the additional factor $(M_\ast/M_\odot)^{-0.25}$ in equation~(\ref{eq:lw2mdot}) and calibrating $f_1=0.115$ (it was a free parameter set to $1.5$ in fig.~5 of \citetalias{SYT}).

\section{Evolution of differential rotation due to shear}
\label{app:diff_rot}
\begin{figure*}
\centering
\includegraphics[width=1.0\textwidth]{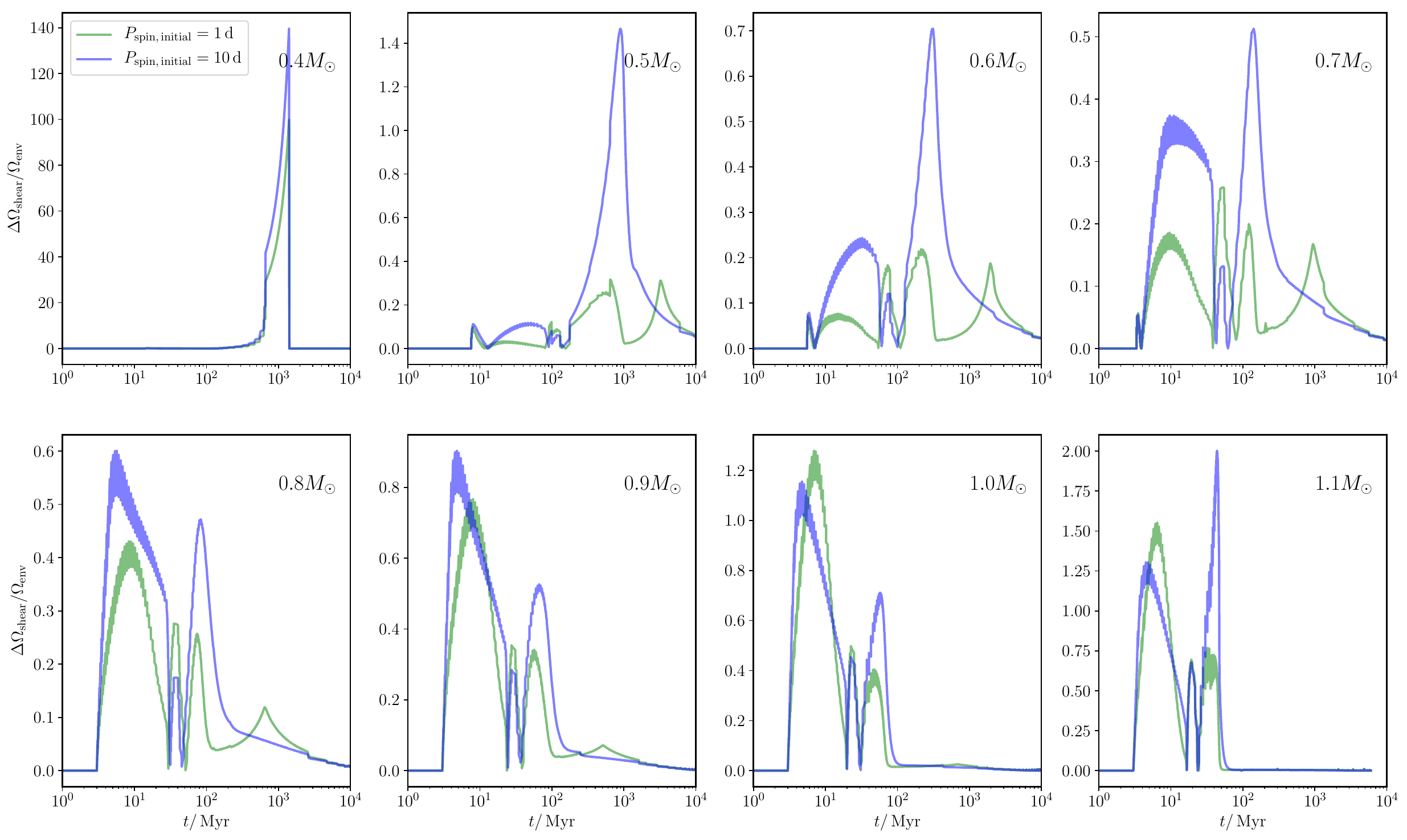}
\caption{The evolution of $\Delta\Omega_\mathrm{shear}/\,\Omega_\mathrm{env}$ with time for various PCDs with $\tau_\mathrm{dl}=3\,$Myr, $f_2=2$, $f_\mathrm{shear}=0.3$ and $\delta=3$. Erratic behaviour is because $\Delta\Omega_\mathrm{shear}/\,\Omega_\mathrm{env}$ is stiff for PCDs at early times before the star transitions from its PMS phase to the MS phase (see Fig.~\ref{fig:holistic_otherobs}).}
\label{fig:holistic_diffrot_analysis}
\end{figure*}

We plot the absolute relative differential rotation at the core-envelope boundary $\Delta\Omega_\mathrm{shear}/\,\Omega_\mathrm{env}$ in Fig.~\ref{fig:holistic_diffrot_analysis}. Owing to rapid changes in the structure of the star, we see some stiffness in $\Delta\Omega_\mathrm{shear}/\,\Omega_\mathrm{env}$ at early times before the star becomes structurally stable. Beyond this, $\Delta\Omega_\mathrm{shear}/\,\Omega_\mathrm{env}$ decreases abruptly. We define this age as $\tau_\mathrm{converge}$ in equation~(\ref{eq:tau_converge}). We see a local maximum in $\Delta\Omega_\mathrm{shear}/\,\Omega_\mathrm{env}$ after $\simeq$ Gyr in fast-rotating K-dwarfs. This can be alleviated by using a larger $f_\mathrm{shear}$ without changing any conclusion significantly (see Fig.~\ref{fig:shear_analysis} for a weak dependence on shear).

\bsp	
\label{lastpage}
\end{document}